\documentclass[12pt]{article}
\usepackage{lineno}
\usepackage{epsfig}
\usepackage{amsmath}
\usepackage{hhline}
\usepackage{amssymb}
\usepackage{times}
\usepackage{cite}
\usepackage{rotating}
\usepackage{url}
\usepackage{captcont}

\newcommand\mct{\mbox{$M_c$}}
\newcommand\mcto{\mbox{$M_c^{\rm opt}$}}

\newcommand{\red}{$\sigma_{\rm red}^{c\bar{c}}\, $}

\newcommand{\dzero}{$D^{0}\, $}

\newlength{\dinwidth}
\newlength{\dinmargin}
\def\Journal#1#2#3#4{{#1} {\bf #2}, (#3) #4}

\def\EPJC{{Eur. Phys. J.} {\bf C}}

\def\be{\begin{equation}}
\def\ee{\end{equation}}
\def\bea{\begin{eqnarray}}
\def\eea{\end{eqnarray}}
\def\etal{{\it et~al.}}

\begin{document}  
\newcommand{\pom}{{I\!\!P}}
\newcommand{\slowpi}{\pi_{\mathit{slow}}}
\newcommand{\fiidiii}{F_2^{D(3)}}
\newcommand{\fiidiiiarg}{\fiidiii\,(\beta,\,Q^2,\,x)}
\newcommand{\n}{1.19\pm 0.06 (stat.) \pm0.07 (syst.)}
\newcommand{\nz}{1.30\pm 0.08 (stat.)^{+0.08}_{-0.14} (syst.)}
\newcommand{\fiidiiiful}{F_2^{D(4)}\,(\beta,\,Q^2,\,x,\,t)}
\newcommand{\fiipom}{\tilde F_2^D}
\newcommand{\ALPHA}{1.10\pm0.03 (stat.) \pm0.04 (syst.)}
\newcommand{\ALPHAZ}{1.15\pm0.04 (stat.)^{+0.04}_{-0.07} (syst.)}
\newcommand{\fiipomarg}{\fiipom\,(\beta,\,Q^2)}
\newcommand{\pomflux}{f_{\pom / p}}
\newcommand{\nxpom}{1.19\pm 0.06 (stat.) \pm0.07 (syst.)}
\newcommand {\gapprox}
   {\raisebox{-0.7ex}{$\stackrel {\textstyle>}{\sim}$}}
\newcommand {\lapprox}
   {\raisebox{-0.7ex}{$\stackrel {\textstyle<}{\sim}$}}
\def\gsim{\,\lower.25ex\hbox{$\scriptstyle\sim$}\kern-1.30ex%
\raise 0.55ex\hbox{$\scriptstyle >$}\,}
\def\lsim{\,\lower.25ex\hbox{$\scriptstyle\sim$}\kern-1.30ex%
\raise 0.55ex\hbox{$\scriptstyle <$}\,}
\newcommand{\pomfluxarg}{f_{\pom / p}\,(x_\pom)}
\newcommand{\dsf}{\mbox{$F_2^{D(3)}$}}
\newcommand{\dsfva}{\mbox{$F_2^{D(3)}(\beta,Q^2,x_{I\!\!P})$}}
\newcommand{\dsfvb}{\mbox{$F_2^{D(3)}(\beta,Q^2,x)$}}
\newcommand{\dsfpom}{$F_2^{I\!\!P}$}
\newcommand{\gap}{\stackrel{>}{\sim}}
\newcommand{\lap}{\stackrel{<}{\sim}}
\newcommand{\fem}{$F_2^{em}$}
\newcommand{\tsnmp}{$\tilde{\sigma}_{NC}(e^{\mp})$}
\newcommand{\tsnm}{$\tilde{\sigma}_{NC}(e^-)$}
\newcommand{\tsnp}{$\tilde{\sigma}_{NC}(e^+)$}
\newcommand{\st}{$\star$}
\newcommand{\sst}{$\star \star$}
\newcommand{\ssst}{$\star \star \star$}
\newcommand{\sssst}{$\star \star \star \star$}

\newcommand{\tw}{\theta_W}
\newcommand{\sw}{\sin{\theta_W}}
\newcommand{\cw}{\cos{\theta_W}}
\newcommand{\sww}{\sin^2{\theta_W}}
\newcommand{\cww}{\cos^2{\theta_W}}
\newcommand{\trm}{m_{\perp}}
\newcommand{\trp}{p_{\perp}}
\newcommand{\trmm}{m_{\perp}^2}
\newcommand{\trpp}{p_{\perp}^2}
\newcommand{\alp}{\alpha_s}

\newcommand{\alps}{\alpha_s}
\newcommand{\sqrts}{$\sqrt{s}$}
\newcommand{\LO}{$O(\alpha_s^0)$}
\newcommand{\Oa}{$O(\alpha_s)$}
\newcommand{\Oaa}{$O(\alpha_s^2)$}
\newcommand{\PT}{p_{\perp}}
\newcommand{\JPSI}{J/\psi}
\newcommand{\sh}{\hat{s}}
\newcommand{\uh}{\hat{u}}
\newcommand{\MP}{m_{J/\psi}}
\newcommand{\PO}{I\!\!P}
\newcommand{\xbj}{x}
\newcommand{\xpom}{x_{\PO}}
\newcommand{\ttbs}{\char'134}
\newcommand{\xpomlo}{3\times10^{-4}}  
\newcommand{\xpomup}{0.05}  
\newcommand{\dgr}{^\circ}
\newcommand{\pbarnt}{\,\mbox{{\rm pb$^{-1}$}}}
\newcommand{\gev}{\,\mbox{GeV}}
\newcommand{\WBoson}{\mbox{$W$}}
\newcommand{\fbarn}{\,\mbox{{\rm fb}}}
\newcommand{\fbarnt}{\,\mbox{{\rm fb$^{-1}$}}}
%
%
\newcommand{\qsq}{\ensuremath{Q^2} }
\newcommand{\gevsq}{\ensuremath{\mathrm{GeV}^2} }
\newcommand{\et}{\ensuremath{E_t^*} }
\newcommand{\rap}{\ensuremath{\eta^*} }
\newcommand{\gp}{\ensuremath{\gamma^*}p }
\newcommand{\dsiget}{\ensuremath{{\rm d}\sigma_{ep}/{\rm d}E_t^*} }
\newcommand{\dsigrap}{\ensuremath{{\rm d}\sigma_{ep}/{\rm d}\eta^*} }

\begin{titlepage}

\noindent
\begin{flushleft}
{\tt DESY 12-172    \hfill    ISSN 0418-9833} \\
{\tt October 2012}                  \\

\end{flushleft}

\vspace{1cm}

\vspace{3.5cm}
\begin{center}
\begin{Large}
\boldmath
{\bf Combination and QCD Analysis of Charm Production Cross Section Measurements in Deep-Inelastic \boldmath $ep$ Scattering at HERA 
}
\unboldmath
\vspace{2cm}

H1 and ZEUS Collaborations

\end{Large}
\end{center}

\vspace{2cm}
\def\gev{\rm GeV}
\def\ie{\it i.e.}
\def\etal{\hbox{$\it et~al.$}}
\def\clb#1 {(#1 Coll.),}
\hyphenation{do-mi-nant
}

\begin{abstract}
\noindent

Measurements of open charm production cross sections in deep-inelastic $ep$ scattering at HERA 
from the H1 and ZEUS Collaborations are combined. Reduced cross sections  \red
for charm production are obtained in the kinematic range of photon virtuality $2.5\le Q^2 \le 2000$ GeV$^2$ 
and Bjorken scaling variable $3\cdot 10^{-5} \le x\le 5\cdot 10^{-2}$. The combination method accounts for the correlations 
of the systematic uncertainties among the different data sets. The combined charm data together 
with the combined inclusive deep-inelastic scattering cross sections from HERA are used as input for 
a detailed  NLO QCD analysis to study the influence of different heavy flavour schemes on 
the parton distribution functions. The optimal values of the charm mass as a parameter in these different 
schemes are obtained. The implications on the NLO predictions for $W^\pm$ and $Z$ production 
cross sections at the LHC are investigated. Using the fixed flavour number scheme, the running mass of 
the charm quark is determined.
\end{abstract}
\vspace{1.5cm}
\begin{center}
Submitted to \EPJC
\end{center}
\end{titlepage}
\newpage

\topmargin-1.cm
\evensidemargin-0.3cm
\oddsidemargin-0.3cm
\textwidth 15.cm
\textheight 680pt
\parindent0.cm
\parskip0.3cm plus0.05cm minus0.05cm
\def\3{\ss}
\newcommand{\address}{ }
\renewcommand{\author}{ }



H.~Abramowicz$^{70, a43}$, 
I.~Abt$^{56}$, 
L.~Adamczyk$^{35}$, 
M.~Adamus$^{83}$, 
R.~Aggarwal$^{14, a11}$, 
C.~Alexa$^{13}$,                
V.~Andreev$^{53}$,             
S.~Antonelli$^{10}$, 
P.~Antonioli$^{9}$, 
A.~Antonov$^{54}$
% ...

\bigskip{\it

   $^{1}$ I. Physikalisches Institut der RWTH, Aachen, Germany \\
 $^{2}$ {\it Institute of Physics and Technology of Ministry of Education and Science of Kazakhstan, Almaty, Kazakhstan}\\
 $^{3}$ {\it NIKHEF and University of Amsterdam, Amsterdam, Netherlands}~$^{b29}$\\
   $^{4}$ Inter-University Institute for High Energies ULB-VUB, Brussels and         Universiteit Antwerpen, Antwerpen, Belgium~$^{b2}$ \\
 $^{5}$ {\it Argonne National Laboratory, Argonne, Illinois 60439-4815, USA}~$^{b13}$\\
   $^{6}$ Vinca Institute of Nuclear Sciences, University of Belgrade,          1100 Belgrade, Serbia \\
 $^{7}$ {\it Andrews University, Berrien Springs, Michigan 49104-0380, USA}\\
   $^{8}$ School of Physics and Astronomy, University of Birmingham,          Birmingham, UK~$^{b16}$ \\
 $^{9}$ {\it INFN Bologna, Bologna, Italy}~$^{b14}$\\
 $^{10}$ {\it University and INFN Bologna, Bologna, Italy}~$^{b14}$\\
 $^{11}$ {\it Physikalisches Institut der Universit\"at Bonn, Bonn, Germany}~$^{b15}$\\
 $^{12}$ {\it H.H.~Wills Physics Laboratory, University of Bristol, Bristol, United Kingdom}~$^{b16}$\\
   $^{13}$ National Institute for Physics and Nuclear Engineering (NIPNE) ,          Bucharest, Romania~$^{b11}$ \\
 $^{14}$ {\it Panjab University, Department of Physics, Chandigarh, India}\\
 $^{15}$ {\it Department of Engineering in Management and Finance, Univ. of the Aegean, Chios, Greece}\\
 $^{16}$ {\it Physics Department, Ohio State University, Columbus, Ohio 43210, USA}~$^{b13}$\\
 $^{17}$ {\it Calabria University, Physics Department and INFN, Cosenza, Italy}~$^{b14}$\\
   $^{18}$ The Henryk Niewodniczanski Institute of Nuclear Physics, Polish Academy of Sciences, Cracow, Poland~$^{b4}$ \\
 $^{19}$ {\it Department of Physics, Jagellonian University, Cracow, Poland}\\
 $^{20}$ {\it Kyungpook National University, Center for High Energy Physics, Daegu, South Korea}~$^{b23}$\\
   $^{21}$ Institut f\"ur Physik, TU Dortmund, Dortmund, Germany~$^{b1}$ \\
   $^{22}$ Joint Institute for Nuclear Research, Dubna, Russia \\
 $^{23}$ {\it INFN Florence, Florence, Italy}~$^{b14}$\\
 $^{24}$ {\it University and INFN Florence, Florence, Italy}~$^{b14}$\\
 $^{25}$ {\it Fakult\"at f\"ur Physik der Universit\"at Freiburg i.Br., Freiburg i.Br., Germany}\\
   $^{26}$ CEA, DSM/Irfu, CE-Saclay, Gif-sur-Yvette, France \\
 $^{27}$ {\it School of Physics and Astronomy, University of Glasgow, Glasgow, United Kingdom}~$^{b16}$\\
$^{28}$ Institut f\"ur Experimentalphysik, Universit\"at Hamburg, Hamburg, Germany~$^{b1}$$^{,}$$^{b21}$\\ 
 $^{29}$ {\it Deutsches Elektronen-Synchrotron DESY, Hamburg, Germany}\\
   $^{30}$ Physikalisches Institut, Universit\"at Heidelberg,          Heidelberg, Germany~$^{b1}$ \\
   $^{31}$ Kirchhoff-Institut f\"ur Physik, Universit\"at Heidelberg,          Heidelberg, Germany~$^{b1}$ \\
   $^{32}$ Max-Planck-Institut f\"ur Kernphysik, Heidelberg, Germany \\
 $^{33}$ {\it Nevis Laboratories, Columbia University, Irvington on Hudson, New York 10027, USA}~$^{b18}$\\
   $^{34}$ Institute of Experimental Physics, Slovak Academy of          Sciences, Ko\v{s}ice, Slovak Republic~$^{b5}$ \\
 $^{35}$ {\it AGH-University of Science and Technology, Faculty of Physics and Applied Computer Science, Krakow, Poland}~$^{b20}$\\
 $^{36}$ {\it Jabatan Fizik, Universiti Malaya, 50603 Kuala Lumpur, Malaysia}~$^{b17}$\\
 $^{37}$ {\it Institute for Universe and Elementary Particles, Chonnam National University, Kwangju, South Korea}\\
 $^{38}$ {\it Institute for Nuclear Research, National Academy of Sciences, Kyiv, Ukraine}\\
 $^{39}$ {\it Department of Nuclear Physics, National Taras Shevchenko University of Kyiv, Kyiv, Ukraine}\\
   $^{40}$ Department of Physics, University of Lancaster,          Lancaster, UK~$^{b16}$ \\
   $^{41}$ Department of Physics, University of Liverpool,         Liverpool, UK~$^{b16}$ \\
   $^{42}$ School of Physics and Astronomy, Queen Mary, University of London,          London, UK~$^{b16}$ \\
 $^{43}$ {\it Imperial College London, High Energy Nuclear Physics Group, London, United Kingdom}~$^{b16}$\\
 $^{44}$ {\it Physics and Astronomy Department, University College London, London, United Kingdom}~$^{b16}$\\
 $^{45}$ {\it Institut de Physique Nucl\'{a13}aire, Universit\'{a13} Catholique de Louvain, Louvain-la-Neuve,\\ Belgium}~$^{b24}$\\
   $^{46}$ Physics Department, University of Lund,          Lund, Sweden~$^{b6}$ \\
   $^{47}$ Departamento de Fisica, CINVESTAV  IPN, M\'exico City, M\'exico~$^{b9}$ \\
 $^{48}$ {\it Department of Physics, University of Wisconsin, Madison, Wisconsin 53706, USA}~$^{b13}$\\
 $^{49}$ {\it Departamento de F\'{\i}sica Te\'orica, Universidad Aut\'onoma de Madrid, Madrid, Spain}~$^{b25}$\\
   $^{50}$ CPPM, Aix-Marseille Univ, CNRS/IN2P3, 13288 Marseille, France \\
 $^{51}$ {\it Department of Physics, McGill University, Montr\'eal, Qu\'ebec, Canada H3A 2T8}~$^{b26}$\\
   $^{52}$ Institute for Theoretical and Experimental Physics, Moscow, Russia~$^{b10}$ \\
   $^{53}$ Lebedev Physical Institute, Moscow, Russia \\
 $^{54}$ {\it Moscow Engineering Physics Institute, Moscow, Russia}~$^{b27}$\\
 $^{55}$ {\it Lomonosov Moscow State University, Skobeltsyn Institute of Nuclear Physics, Moscow, Russia}~$^{b28}$\\
 $^{56}$ {\it Max-Planck-Institut f\"ur Physik, M\"unchen, Germany}\\
   $^{57}$ LAL, Universit\'e Paris-Sud, CNRS/IN2P3, Orsay, France \\
 $^{58}$ {\it Department of Physics, University of Oxford, Oxford, United Kingdom}~$^{b16}$\\
   $^{59}$ STFC, Rutherford Appleton Laboratory, Didcot, Oxfordshire, UK~$^{b16}$ \\
 $^{60}$ {\it INFN Padova, Padova, Italy}~$^{b14}$\\
 $^{61}$ {\it Dipartimento di Fisica dell' Universit\`a and INFN, Padova, Italy}~$^{b14}$\\
   $^{62}$ LLR, Ecole Polytechnique, CNRS/IN2P3, Palaiseau, France \\
   $^{63}$ Faculty of Science University of Montenegro, Podgorica, Montenegro~$^{b12}$ \\
   $^{64}$ Institute of Physics of the Academy of Sciences of the Czech Republic,  Praha, Czech Republic~$^{b7}$ \\
   $^{65}$ Faculty of Mathematics and Physics of Charles University, Praha, Czech Republic~$^{b7}$ \\
 $^{66}$ {\it Department of Particle Physics and Astrophysics, Weizmann Institute, Rehovot, Israel}\\
   $^{67}$ Dipartimento di Fisica Universit\`a di Roma Tre   and INFN Roma~3, Roma, Italy \\
 $^{68}$ {\it Dipartimento di Fisica, Universit\`a 'La Sapienza' and INFN, Rome, Italy}~$^{b14}$\\
   $^{69}$ Institute for Nuclear Research and Nuclear Energy, Sofia, Bulgaria \\
 $^{70}$ {\it Raymond and Beverly Sackler Faculty of Exact Sciences, School of Physics, Tel Aviv University, Tel Aviv, Israel}~$^{b30}$\\
 $^{71}$ {\it Polytechnic University, Tokyo, Japan}~$^{b22}$\\
 $^{72}$ {\it Department of Physics, Tokyo Institute of Technology, Tokyo, Japan}~$^{b22}$\\
 $^{73}$ {\it Department of Physics, University of Tokyo, Tokyo, Japan}~$^{b22}$\\
 $^{74}$ {\it Tokyo Metropolitan University, Department of Physics, Tokyo, Japan}~$^{b22}$\\
 $^{75}$ {\it Universit\`a di Torino and INFN, Torino, Italy}~$^{b14}$\\
 $^{76}$ {\it Universit\`a del Piemonte Orientale, Novara, and INFN, Torino, Italy}~$^{b14}$\\
 $^{77}$ {\it Department of Physics, University of Toronto, Toronto, Ontario, Canada M5S 1A7}~$^{b26}$\\
 $^{78}$ {\it Institute of Particle and Nuclear Studies, KEK, Tsukuba, Japan}~$^{b22}$\\
   $^{79}$ Institute of Physics and Technology of the Mongolian  Academy of Sciences, Ulaanbaatar, Mongolia \\
 $^{80}$ {\it Department of Physics, Pennsylvania State University, University Park, Pennsylvania 16802, USA}~$^{b18}$\\
   $^{81}$ Paul Scherrer Institut,    Villigen, Switzerland \\
 $^{82}$ {\it Faculty of Physics, University of Warsaw, Warsaw, Poland}\\
 $^{83}$ {\it National Centre for Nuclear Research, Warsaw, Poland}\\
   $^{84}$ Fachbereich C, Universit\"at Wuppertal, Wuppertal, Germany \\
   $^{85}$ Yerevan Physics Institute, Yerevan, Armenia \\
 $^{86}$ {\it Meiji Gakuin University, Faculty of General Education, Yokohama, Japan}~$^{b22}$\\
 $^{87}$ {\it Department of Physics, York University, Ontario, Canada M3J 1P3}~$^{b26}$\\
   $^{88}$ Departamento de Fisica Aplicada,        CINVESTAV, M\'erida, Yucat\'an, M\'exico~$^{b9}$ \\
 $^{89}$ {\it Deutsches Elektronen-Synchrotron DESY, Zeuthen, Germany}\\
   $^{90}$ Institut f\"ur Teilchenphysik, ETH, Z\"urich, Switzerland~$^{b8}$ \\
   $^{91}$ Physik-Institut der Universit\"at Z\"urich, Z\"urich, Switzerland~$^{b8}$ \\

\bigskip
 $ ^{a1}$ Also at Rechenzentrum, Universit\"at Wuppertal,
          Wuppertal, Germany \\
 $ ^{a2}$ Also at IPNL, Universit\'e Claude Bernard Lyon 1, CNRS/IN2P3,
          Villeurbanne, France \\
 $ ^{a3}$ Also at CERN, Geneva, Switzerland \\
 $ ^{a4}$ Also at Ulaanbaatar University, Ulaanbaatar, Mongolia \\
 $ ^{a5}$ Supported by the Initiative and Networking Fund of the
          Helmholtz Association (HGF) under the contract VH-NG-401. \\
 $ ^{a6}$ Absent on leave from NIPNE-HH, Bucharest, Romania \\
 $ ^{a7}$ Also at  Department of Physics, University of Toronto,
          Toronto, Ontario, Canada M5S 1A7 \\
 $ ^{a8}$ Also at LAPP, Universit\'e de Savoie, CNRS/IN2P3,
          Annecy-le-Vieux, France \\
$^{a9}$ Now at University of Salerno, Italy\\
$^{a10}$ Now at Queen Mary University of London, United Kingdom\\
$^{a11}$ Also funded by Max Planck Institute for Physics, Munich, Germany\\
$^{a12}$ Also Senior Alexander von Humboldt Research Fellow at Hamburg University, Institute of Experimental Physics, Hamburg, Germany\\
$^{a13}$ Also at Cracow University of Technology, Faculty of Physics, Mathemathics and Applied Computer Science, Poland\\
$^{a14}$ Supported by the research grant No. 1 P03B 04529 (2005-2008)\\
$^{a15}$ Supported by the Polish National Science Centre, project No. DEC-2011/01/BST2/03643\\
$^{a16}$ Now at Rockefeller University, New York, NY 10065, USA\\
$^{a17}$ Now at DESY group FS-CFEL-1\\
$^{a18}$ Now at Institute of High Energy Physics, Beijing, China\\
$^{a19}$ Now at DESY group FEB, Hamburg, Germany\\
$^{a20}$ Also at Moscow State University, Russia\\
$^{a21}$ Now at University of Liverpool, United Kingdom\\
$^{a22}$ Now at CERN, Geneva, Switzerland\\
$^{a23}$ Also affiliated with Universtiy College London, UK\\
$^{a24}$ Now at Goldman Sachs, London, UK\\
$^{a25}$ Also at Institute of Theoretical and Experimental Physics, Moscow, Russia\\
$^{a26}$ Also at FPACS, AGH-UST, Cracow, Poland\\
$^{a27}$ Partially supported by Warsaw University, Poland\\
$^{a28}$ Now at Istituto Nucleare di Fisica Nazionale (INFN), Pisa, Italy\\
$^{a29}$ Now at Haase Energie Technik AG, Neum\"unster, Germany\\
$^{a30}$ Now at Department of Physics, University of Bonn, Germany\\
$^{a31}$ Also affiliated with DESY, Germany\\
$^{a32}$ Also at University of Tokyo, Japan\\
$^{a33}$ Now at Kobe University, Japan\\
$^{a35}$ Supported by DESY, Germany\\
$^{a36}$ Member of National Technical University of Ukraine, Kyiv Polytechnic Institute,  Kyiv, Ukraine\\
$^{a37}$ Member of National University of Kyiv - Mohyla Academy, Kyiv, Ukraine\\
$^{a38}$ Partly supported by the Russian Foundation for Basic Research, grant 11-02-91345-DFG\_a\\
$^{a39}$ Alexander von Humboldt Professor; also at DESY and University of Oxford\\
$^{a40}$ STFC Advanced Fellow\\
$^{a41}$ Now at LNF, Frascati, Italy\\
$^{a42}$ This material was based on work supported by the National Science Foundation, while working at the Foundation.\\
$^{a43}$ Also at Max Planck Institute for Physics, Munich, Germany, External Scientific Member\\
$^{a44}$ Now at Tokyo Metropolitan University, Japan\\
$^{a45}$ Now at Nihon Institute of Medical Science, Japan\\
$^{a46}$ Now at Osaka University, Osaka, Japan\\
$^{a47}$ Also at Lodz University, Poland\\ 
$^{a48}$ Member of Lodz  University, Poland\\ 
$^{a49}$ Now at Department of Physics, Stockholm University, Stockholm, Sweden\\
$^{a50}$ Also at Cardinal Stefan Wyszy\'nski University, Warsaw, Poland\\
\bigskip

 $^{b1}$ Supported by the Bundesministerium f\"ur Bildung und Forschung, FRG, under contract numbers 05H09GUF, 05H09VHC, 05H09VHF,  05H16PEA \\
 $^{b2}$ Supported by FNRS-FWO-Vlaanderen, IISN-IIKW and IWT  and  by Interuniversity Attraction Poles Programme,  Belgian Science Policy \\
 $^{b4}$ Supported by Polish Ministry of Science and Higher  Education, grants  DPN/N168/DESY/2009 and DPN/N188/DESY/2009 \\
 $^{b5}$ Supported by VEGA SR grant no. 2/7062/ 27 \\
 $^{b6}$ Supported by the Swedish Natural Science Research Council \\
 $^{b7}$ Supported by the Ministry of Education of the Czech Republic under the projects  LC527, INGO-LA09042 and   MSM0021620859 \\
 $^{b8}$ Supported by the Swiss National Science Foundation \\
 $^{b9}$ Supported by  CONACYT,  M\'exico, grant 48778-F \\
 $^{b10}$ Russian Foundation for Basic Research (RFBR), grant no 1329.2008.2 and Rosatom \\
 $^{b11}$ Supported by the Romanian National Authority for Scientific Research  under the contract PN 09370101 \\
 $^{b12}$ Partially Supported by Ministry of Science of Montenegro,  no. 05-1/3-3352 \\
$^{b13}$  Supported by the US Department of Energy\\
$^{b14}$  Supported by the Italian National Institute for Nuclear Physics (INFN) \\
$^{b15}$  Supported by the German Federal Ministry for Education and Research (BMBF), under contract No. 05 H09PDF\\
$^{b16}$  Supported by the Science and Technology Facilities Council, UK\\
$^{b17}$  Supported by HIR and UMRG grants from Universiti Malaya, and an ERGS grant from the Malaysian Ministry for Higher Education\\
$^{b18}$  Supported by the US National Science Foundation. Any opinion, findings and conclusions or recommendations expressed in this material are those of the authors and do not necessarily reflect the views of the  National Science Foundation.\\
$^{b20}$  Supported by the Polish Ministry of Science and Higher Education and its grants for Scientific Research\\
$^{b21}$  Supported by the German Federal Ministry for Education and Research (BMBF), under contract No. 05h09GUF, and the SFB 676 of the Deutsche Forschungsgemeinschaft (DFG) \\
$^{b22}$  Supported by the Japanese Ministry of Education, Culture, Sports, Science and Technology (MEXT) and its grants for Scientific Research\\
$^{b23}$  Supported by the Korean Ministry of Education and Korea Science and Engineering Foundation\\
$^{b24}$  Supported by FNRS and its associated funds (IISN and FRIA) and by an Inter-University Attraction Poles Programme subsidised by the Belgian Federal Science Policy Office\\
$^{b25}$  Supported by the Spanish Ministry of Education and Science through funds provided by CICYT\\
$^{b26}$  Supported by the Natural Sciences and Engineering Research Council of Canada (NSERC) \\
$^{b27}$  Partially supported by the German Federal Ministry for Education and Research (BMBF)\\
$^{b28}$  Supported by RF Presidential grant N 3920.2012.2 for Leading Scientific Schools and by the Russian Ministry of Education and Science through its grant for Scientific Research on High Energy Physics \\
$^{b29}$  Supported by the Netherlands Foundation for Research on Matter (FOM)\\
$^{b30}$  Supported by the Israel Science Foundation\\

\vspace{0.2cm}
$^{\dagger}$    deceased \\

}


\newpage
\section{Introduction}
Measurements of open charm production in deep-inelastic electron\footnote{In this paper `electron' is used to denote both electron and positron if not otherwise stated. }-proton scattering (DIS) at HERA provide important input for stringent tests of the theory of strong interactions, quantum chromodynamics ({QCD}).
Previous measurements \cite{h196,zeusdstar97,h1gluon,zd97,h1f2c,zd00,h1dmesons,h1vertex05,h1ltt_hera1,h1dstar_hera1,zeusdmesons,zd0dp,zmu,h1ltt_hera2,h1dstarhighQ2,zeusdpluslambda,h1cbjets,h1dstar_hera2} have demonstrated that charm quarks are predominantly produced by the boson-gluon-fusion process, $\gamma g\rightarrow c\overline{c}$, which is sensitive to the gluon distribution in the proton.

The mass of the charm quark, $m_c$, provides a sufficiently high scale necessary to apply perturbative QCD (pQCD). 
However, additional scales are involved in charm production, 
e.g. the virtuality, $Q^2$, of the exchanged photon in case of DIS and the transverse momenta, $p_T$, of the outgoing quarks. The presence of several hard scales complicates the QCD calculations for charm production.
Depending on the details of the treatment of $m_c$, $Q$ and $p_T$, different approaches in pQCD have been formulated.
In this paper, the massive fixed-flavour-number-scheme (FFNS) \cite{riemersma,gjr,ct10f3,mstw08f3,abm11} and different implementations of the variable-flavour-number-scheme (VFNS) \cite{zmvfns,acot,bmsn,s_acot_chi,rt_std,FONLL,fonllb_and_c,rt_opt} are considered.

At HERA different techniques have been used 
to measure open charm production cross sections in DIS.
The full reconstruction of $D$ or $D^*$ mesons~\cite{
h196,zeusdstar97,zd97,h1f2c,zd00,h1dstar_hera1,zeusdmesons,zd0dp,h1dstarhighQ2,h1dstar_hera2}, the long lifetime of heavy 
flavoured hadrons \cite{h1dmesons,h1vertex05,h1ltt_hera1,zd0dp,h1ltt_hera2} or their semi-leptonic decays~\cite{zmu} are 
exploited.  
In general, the best signal-to-background ratio of the charm samples 
is observed in the analysis of fully reconstructed $D^*$ mesons. 
However, the branching ratios are small and the phase space of charm production accessible with $D^*$ mesons is restricted considerably
because all products from the $D^*$ meson decay have to be measured.
The usage of semi-leptonic decays of charmed hadrons for the analysis of charm production can profit from large branching fractions and a better coverage in polar angle at the cost of a worse signal-to-background ratio.
Fully inclusive analyses using lifetime information are not hampered by specific branching ratios and are in addition sensitive to low transverse momenta.
Among the methods used it has the largest phase space coverage, however 
it yields the worst signal-to-background ratio.

In this paper the published data of H1 \cite{h1ltt_hera1,h1dstar_hera1,h1ltt_hera2,h1dstarhighQ2,h1dstar_hera2} and ZEUS \cite{zd97,zd00,zd0dp,zmu} are combined.
All 
publications on data sets\footnote{
The data taken up to the year 2000 and data taken after 2002 are referred to as HERA-I and HERA-II, respectively.} 
are included 
for which the necessary information on systematic uncertainties needed for the combination is available 
and which have not been superseded. 
For the combination, the published cross sections in the restricted phase space regions
of the individual measurements are extrapolated to the full phase space of charm production in a coherent manner by the use of FFNS calculations in next-to-leading order (NLO). 
This includes the coherent treatment of the related  systematic uncertainties. 

 The combination is based on the procedure described in \cite{glazov,H1comb,DIScomb}.  
The correlated systematic uncertainties and the normalisation of the different measurements are accounted for 
such that one consistent data set is obtained. Since different experimental techniques of charm tagging 
have been employed using different detectors and methods of kinematic reconstruction, this combination 
leads to a significant reduction of statistical and systematic uncertainties.

The combined charm data are used together with the combined inclusive DIS cross sections \cite{DIScomb} to perform a detailed QCD 
analysis using different models of charm production in DIS. The r\^ole of the value for the charm quark mass which enters as a parameter in these models is investigated and the optimal value of the charm quark mass parameter is determined for each of the QCD calculations considered. The impact of this optimisation on predictions of $W^\pm$ and $Z$ 
production cross sections at the LHC is discussed. 
The running mass of the charm quark is determined using the modified minimal subtraction scheme (${\overline{\text{MS}}}$) variant \cite{ABKMMSbar} of the FFNS.
 
The paper is organised as follows. In section \ref{sect:theory} the different theoretical schemes of charm 
production are briefly reviewed. The data samples used for the combination and the details of the combination 
procedure are described in section \ref{Sect:combination}. The results on the combined reduced cross section 
are presented in section \ref{sect:results}. 
The predictions from different QCD approaches for charm production in DIS are compared to the measurement in section \ref{sect:comparisons}. The QCD analysis is presented in section \ref{sect:qcdana}. Conclusions are given in section \ref{sect:conclude}.
\section{Open charm production in DIS}
\label{sect:theory}
In this paper, charm production via neutral-current deep-inelastic $ep$ scattering is considered.
In the kinematic domain addressed, where the virtuality $Q^2$ of the exchanged boson is small, $Q^2 \ll M_Z^2$, 
charm production is dominated by virtual photon exchange.  
The cross section may then be written in terms of the structure functions $F^{c\bar c}_2(x,Q^2)$ and 
$F^{c\bar c}_L(x,Q^2)$ as
\begin{equation}
\frac{{\rm d}^2 \sigma^{c\bar{c}}}{{\rm d}x {\rm d}Q^2} = \frac{2\pi \alpha^2(Q^2)}{xQ^4} ( [1+(1-y)^2]F^{c\bar c}_2 (x,Q^2) -y^2F_L^{c\bar c} (x,Q^2) ).
\end{equation}
Here $x=Q^2/2p\cdot q\,$ is the Bjorken scaling variable and $y=p\cdot q/p\cdot l\,$ is the inelasticity  with $p$, $q$ and $l$ 
denoting the 4-momenta of the proton, photon and electron, respectively, and $Q^2=-q^2$. The 
suffix $c\bar c$ indicates the presence of a $c\bar c$ pair in the final state,
including all possible QCD production processes. The cross
section ${{\rm d}^2 \sigma^{c\bar{c}}}/{{\rm d}x {\rm d}Q^2}$ is given at the
Born level without QED and electro-weak radiative corrections, except for the running electromagnetic coupling, $\alpha(Q^2)$.

In this paper, the results are presented in terms of reduced cross sections, defined as follows:
\begin{eqnarray}
\sigma_{\rm red}^{c\bar{c}}&=&\frac{{\rm d}^2\sigma^{c\bar{c}}}{{\rm d}x {\rm d}Q^2} \cdot \frac{xQ^4}{2\pi\alpha^2(Q^2)\,(1+(1-y)^2)}\cr
&=&F_2^{c\bar{c}}-\frac{y^2}{1+(1-y)^2}F^{c\bar{c}}_L.
\end{eqnarray}
The contribution $F_L^{c\bar{c}}$, originating from the exchange of longitudinally polarised photons, is small in the kinematic range of this 
analysis and reaches up to a few per cent only at high $y$~\cite{daum_fl}.

The above definition of $F^{c\bar c}_{2 (L)} (x,Q^2)$ (also denoted as $\tilde F_c$ \cite{FONLL} or
$F_{c,SI}$ \cite{CTEQNNLO}) is suited for 
measurements in which charm is explicitly detected. It differs 
from what is sometimes used in theoretical calculations in which
$F^c_{2 (L)} (x,Q^2)$ \cite{FONLL,rt_std,MSTWF2} is defined as the contribution to the inclusive $F_{2(L)}
(x,Q^2)$ in which the virtual photon couples directly to a $c$ or $\bar{c}$
quark. The latter excludes contributions from final state gluon splitting to 
a $c\bar c$ pair in events where the photon couples directly to a light quark, 
and contributions from events in which the photon is replaced by a gluon
from a hadron-like resolved photon.
As shown in table 1 of ~\cite{FONLL}, 
the gluon splitting contribution is expected to be 
small enough to allow
a reasonable comparison of the experimental results 
to theoretical predictions using this definition.
The hadron-like resolved photon contribution is expected to be heavily 
suppressed at high $Q^2$, but might not be completely negligible in the low $Q^2$ region.
From the point of view of pQCD it appears at $O(\alpha_s^3)$ and it is
neglected in all theoretical calculations used in this paper. 

At photon virtualities not much larger than the charm quark mass, charm production in DIS is described 
in the framework of pQCD by flavour creation through the virtual photon-gluon-fusion process.
Since a $c\bar c$ pair is being produced, there is a natural lower
cutoff of $2m_c$ for the mass of the hadronic final state.
The non-zero mass influences the kinematics and higher order 
corrections in essentially all the HERA phase space.
Therefore the correct treatment of the mass of charm and beauty quarks is of 
particular importance in the QCD analysis and 
determination of parton distribution functions (PDFs) of the proton.
In the following, the different approaches 
used in the treatment of the charm quark mass in pQCD calculations are discussed.

\subsection{Zero Mass Variable Flavour Number Scheme} 
In the zero-mass variable-flavour-number-scheme ({ZM-VFNS}) \cite{zmvfns} the charm quark mass is set to zero in
the computation of the matrix elements and kinematics, and a 
threshold is introduced
at $Q^2 \sim m_c^2$, below which the charm production cross section is assumed to vanish. 
The charm quark is also excluded from the parton evolution and only three light flavours are left active. 
Above this threshold, charm is treated as a massless parton in the
proton, leading to the introduction of the charm quark distribution function of the proton.
The transition from three to four active flavours in the parton evolution follows the BMSN prescription~\cite{bmsn}.
The lowest order process for charm production in this approach is
the quark-parton-model like scattering at order zero in $\alpha_s$.
The running of $\alpha_s$ is calculated using three flavours ($u,d,s$) below 
the scale $m_c$, and using four or five flavours (including charm and beauty) 
above the respective threshold scales.
The main advantage of this scheme is that the $Q^2$ evolution of the
charm density provides a resummation of  
terms proportional to $\log(Q^2/m_c^2)$ that may be large at large $Q^2$.
It has been shown \cite{h1dstarhighQ2,h1dstar_hera2} that
this approach does not describe the
charm production data at HERA.

\subsection{Fixed Flavour Number Scheme} 

In the fixed-flavour-number-scheme (FFNS)
the charm quark is treated as massive at all scales, and is not considered as a parton in the proton. 
The number of active flavours, $n_f$, is fixed to three, and charm quarks are 
assumed to be produced only in the 
hard scattering process. Thus the leading order (LO) process for charm
production is the boson-gluon-fusion process at $O(\alpha_s)$.
The next-to-leading order (NLO) coefficient functions for 
charm production at $O(\alpha_s^2)$ in the FFNS 
were calculated in~\cite{riemersma} and adopted by many global QCD analysis 
groups~\cite{abm11,gjr,ct10f3,mstw08f3}, providing PDFs in the FFNS. 
In the data analysis presented in this paper, the prediction of open charm 
production in the FFNS at NLO is used to calculate 
inclusive~\cite{riemersma} and exclusive~\cite{hvqdis} quantities. 
   
In the calculations~\cite{riemersma,hvqdis} the pole mass definition~\cite{pdg} is used 
for the charm quark mass, and gluon splitting contributions are included. 
In a recent variant of the FFNS scheme (ABM FFNS)~\cite{ABKMMSbar}, 
the running 
mass definition in the modified minimal subtraction scheme ($\overline{{\text {MS}}}$) is used 
instead. This scheme has the advantage of reducing the sensitivity of 
the cross sections to higher order corrections, and improving the theoretical 
precision of the mass definition.

To $O(\alpha_s)$, 
which is relevant for the calculation of cross sections to $O(\alpha_s^2)$, the  
$\overline{\text {MS}}$ and pole masses are related by~\cite{mc_run}
\begin{equation} \label{eq:mrunpol}
m_c(Q) = m_{c,{\mathrm{pole}}} \left[1 - \frac{\alpha_s}{\pi} 
   - \frac{3\alpha_s}{4\pi} \ln\left(\frac{Q^2}{m_c(m_c)^2}\right)\right],
\end{equation}
i.e. the running mass evaluated at the scale $Q=m_c$ is 
smaller than the pole mass.  

\subsection{General Mass Variable Flavour Number Scheme} 

In the general-mass variable-flavour-number-schemes (GM-VFNS)
charm production is treated in the FFNS
in the low $Q^2$ region, where the mass effects are largest, and in the ZM-VFNS
approach at high $Q^2$, where the effect of resummation
is most noticeable. At intermediate scales an interpolation 
is made between the FFNS and the ZM-VFNS, avoiding double counting of 
common terms. 
This scheme is expected to combine the advantages of the FFNS
and ZM-VFNS, while introducing 
some level of arbitrariness in the treatment of the interpolation. Different implementations 
of the GM-VFNS are available~\cite{FONLL,rt_std,bmsn,acot,s_acot_chi,rt_opt,fonllb_and_c} and are used by the global QCD analysis groups. 

The freedom introduced by choosing an interpolation approach prevents
a clear interpretation of the charm mass in terms of a specific renormalisation scheme. 
Therefore 
the charm mass appearing in the {GM-VFNS}
will be treated in the following sections as an effective mass parameter, \mct, of 
the individual interpolation models.
\section{Combination of H1 and ZEUS measurements}
\label{Sect:combination}
\subsection{Data samples}
\label{sect:samples}
The H1\cite{h1} and ZEUS~\cite{zeus} detectors were general purpose
instruments which consisted of tracking systems surrounded by
electromagnetic and hadronic calorimeters and muon detectors, ensuring
close to $4\pi$ coverage of the $ep$ interaction point. Both detectors
were equipped with high-resolution silicon vertex detectors: the
Central Silicon Tracker \cite{H1cst} for H1 and the Micro Vertex Detector \cite{ZEUSmvd} for ZEUS. 

The data sets included in the combination are listed in table~\ref{t:sets} and correspond 
to $155$ different cross section measurements. The combination includes measurements of charm production performed using
different tagging techniques: 
the reconstruction of particular 
decays of $D$-mesons ~\cite{zd97,zd00,h1dstar_hera1,zd0dp,h1dstarhighQ2,h1dstar_hera2}, 
the inclusive analysis of 
tracks exploiting lifetime information~\cite{h1ltt_hera2} or 
the reconstruction of muons from 
charm semi-leptonic decays~\cite{zmu}. 

The results of the inclusive lifetime analysis~\cite{h1ltt_hera2} are directly taken from the 
original measurement in the form of \red. In the case of $D$-meson and muon measurements, the inputs to the
combination are visible cross sections $\sigma_{\rm vis, bin}$ defined as the $D$ (or $\mu$) production cross 
section in a particular $p_T$ and $\eta$ range, reported in the corresponding publications\footnote{A 
misprint was found in table 3 of~\cite{zd00}: for the rows 22 and 23 the $y$ ranges should read $0.22-0.10$ 
and $0.10-0.02$, respectively. Another misprint was found in table 2 of~\cite{zmu}: the $Q^2$ range in the last row 
should be $400:10000$ GeV$^2$.}, in bins of $Q^2$ and $y$ or $x$. Where necessary, the 
beauty contribution to the inclusive cross sections of $D$ meson production is subtracted 
using the estimates of the corresponding papers. The measured cross 
sections include corrections for radiation of a real photon from the incoming or outgoing lepton and for virtual 
electroweak effects using the HERACLES program~\cite{heracles}. QED corrections to the incoming 
and outgoing quarks were neglected. All $D$-meson cross sections are updated using the most recent branching ratios~\cite{pdg}.

\subsection{Extraction of \boldmath \red  \unboldmath from visible cross sections}
\label{sect:extract}
In the case of $D$-meson and muon production, \red is obtained from the visible
cross sections $\sigma_{\rm vis, bin}$ measured in a limited phase space using a common theory.
The reduced charm cross section at a reference ($x, Q^2$) point is extracted  according to 
\begin{equation}
\sigma_{\rm red}^{\rm c\bar{c}}(x,Q^2)=\sigma_{\rm vis, bin}
\frac{\sigma_{\rm red}^{\rm c\bar{c}, th}(x,Q^2)}{\sigma^{\rm th}_{\rm vis, bin}}.\label{formel_ftc}
\end{equation}
The program from Riemersma et al.~\cite{riemersma} and the program HVQDIS~\cite{hvqdis}
are used to calculate, in NLO FFNS, the reduced cross sections 
 $\sigma_{\rm red}^{\rm c\bar{c}, th}(x,Q^2)$ 
 and the visible cross sections
$\sigma^{\rm th}_{\rm vis, bin}$, respectively.
The following parameters are used consistently in both NLO calculations and the corresponding variations are 
used to estimate the associated uncertainties on the extraction of \red:
\begin{itemize}
 \item {\bf pole mass of the charm quark} $m_c=1.5 \pm 0.15$ GeV;
 \item {\bf renormalisation and factorisation scales}
  $\mu_f=\mu_r=\sqrt{Q^2+4m_c^2}$, varied simultaneously up or down by
  a factor of two;
 \item {\bf strong coupling constant} $\alpha_s^{n_f=3}(M_Z) = 0.105 \pm 0.002$,
 corresponding to $\alpha_s^{n_f=5}(M_Z) = 0.116 \pm 0.002$;
\item {\bf the proton structure} is described by a series of {FFNS} variants of the HERAPDF1.0 set~\cite{DIScomb} 
at NLO, evaluated for $m_c=1.5\pm0.15$~GeV and for  $\alpha_s^{n_f=3}(M_Z) = 0.105 \pm 0.002$. 
For the light flavour contribution,  the renormalisation and factorisation scales are set to 
$\mu_r=\mu_f=Q$, 
while for the heavy quark contributions the scales of 
$\mu_f=\mu_r=\sqrt{Q^2+4m_{Q}^2}$
 are used, 
with $m_{Q}$ being the mass of the charm or beauty quark. Additional PDF sets are evaluated, in which the 
scales are varied simultaneously by a factor of two up or down. Only the scale variation in the heavy 
quark contribution has a sizeable effect on the PDFs. The experimental, model and parameterisation uncertainties of the PDFs
at $68\%$ C.L. are also included in the determination of the PDF uncertainties on \red.
For estimating the uncertainties of the NLO calculations~\cite{riemersma,hvqdis} due to the respective
choice of the scales, $\alpha_s$ and $m_c$, the appropriate PDF set is used. 
The effects of the PDF uncertainties are calculated according to the
HERAPDF1.0  prescription~\cite{DIScomb}.
\end{itemize}

The cross sections $\sigma_{\rm vis, bin}^{\rm th}$ 
depend, in addition to the kinematics of the charm quark production mechanism, also on the 
fragmentation of the charm quark into particular hadrons. 
The charm quark fragmentation function has been measured by H1~\cite{h1frag} and ZEUS~\cite{zeusdmesons} 
using the production of $D^*$ mesons, with and without associated jets, in DIS and 
photoproduction ($Q^2\approx 0$~GeV$^2$). 
In the calculation of $\sigma_{\rm vis, bin}^{\rm th}$ the fragmentation is performed in the $\gamma^*$-$p$ centre-of-mass frame, 
using for the fraction of the charm quark momentum carried by the
charmed meson a fragmentation function 
which is controlled by a single parameter, $\alpha_K$~\cite{kart}. 
The parameter relevant for charm fragmentation into $D^*$ mesons
has been determined~\cite{h1frag, zeusdmesons} for the NLO FFNS calculation for
three different kinematic and jet requirements,
which correspond approximately to three different regions of 
the $\gamma^*$-parton centre-of-mass energy squared, $\hat{s}$.
The values of $\alpha_K$, together with the corresponding
ranges in $\hat{s}$,  are listed in table~\ref{t:kart}.
The fragmentation is observed to become softer with increasing $\hat{s}$, as
expected from the evolution of the fragmentation function.
The limits on the $\hat{s}$ ranges
are determined with HVQDIS by applying the jet requirements of the
individual analysis on parton level.  The  $\alpha_K$ parameters and the
$\hat{s}$ ranges were varied according to their uncertainties  to
evaluate the corresponding uncertainty on  $\sigma_{\rm vis, bin}^{\rm th}$.

Since ground-state $D$ mesons partly originate from decays of $D^*$ and other excited
mesons, the corresponding charm fragmentation function is softer than
that measured using $D^*$ mesons.
From kinematic considerations~\cite{cacciari}, supported 
by experimental measurements~\cite{belle}, the expectation value for
the fragmentation function of charm into $D^{0, {\rm no}D^{*+}}$,
$D^+$ and in the mix of charm hadrons decaying into muons, has to be
reduced by $\approx 5\%$ with respect to that for $D^*$ mesons. The
values of $\alpha_K$ for the fragmentation into ground state hadrons, used for the $D^{0, {\rm no}D^{*+}}$, $D^+$ and $\mu$ measurements, have been 
re-evaluated accordingly and are reported in table~\ref{t:kart}.  

Transverse fragmentation is simulated assigning to charmed hadrons a
transverse momentum, $k_T$, with respect 
to the charm quark direction, according to $f(k_T) = k_T \exp(-2 k_T/\langle k_T\rangle)$. The average $\langle k_T\rangle$ is set
to $0.35\pm 0.15$~GeV~\cite{ptkink}.

The fragmentation fractions of charm quarks into specific $D$ mesons are listed 
in table~\ref{t:ff}. They are obtained from the average of $e^+e^-$ and $ep$ results~\cite{lohrmann}.
The semi-leptonic branching fraction $B(c\rightarrow \mu)$~\cite{pdg} is also given. 
The decay spectrum of leptons from charm decays is taken from~\cite{cleo}.

To evaluate the extrapolation uncertainty on the extracted reduced cross section, \red, all the above parameters are varied by the quoted  
uncertainties and each variation is considered as a correlated uncertainty among the measurements to which it applies.
The dominant contributions arise from the variation of the fragmentation function (average $3-5\%$) and from the
variation of the renormalisation and factorisation scales (average $5-6\%$, reaching $15\%$ at lowest $Q^2$).
In a few cases, the symmetric variation of model parameters results in an asymmetric uncertainty on the cross section. In 
such cases, the larger 
difference with respect to the default cross section is applied symmetrically as systematic uncertainty.

\subsection{Common $\boldsymbol{x-Q^2}$ grid}

Except for the H1 lifetime analysis \cite{h1ltt_hera2}, the values of \red for individual measurements are determined 
at the $52$ ($x, Q^2$) points of a common grid. The grid points are chosen such that they are close to the
centre-of gravity in $x$ and $Q^2$ of the corresponding $\sigma_{\rm vis,bin}$ bins, taking advantage of the fact that 
the binnings used by the H1 and ZEUS experiments are similar.
Prior to the combination, the H1 lifetime analysis measurements are transformed, when needed,
to the common grid ($x, Q^2$) points using the NLO FFNS calculation~\cite{riemersma}.
The resulting scaling factors are always smaller than $18\%$ and
the associated uncertainties, obtained by varying the charm mass, the scales and the PDFs, are negligible.
For all but five grid points at least two measurements enter into the combination. 

\subsection{Combination method}
\label{sect:combine}
The combination of the data sets uses the $\chi^2$ minimisation method developed for the 
combination of inclusive DIS cross sections~\cite{glazov,DIScomb}. 
The $\chi^2$ function takes into account the correlated 
systematic uncertainties for the H1 and ZEUS cross section measurements.
For an individual data set, $e$, the $\chi^2$~function is defined as
\begin{equation}
\chi^2_{{\rm exp},e}\left(\boldsymbol{m},\boldsymbol{b}\right) = 
 \sum_i
 \frac{\left(m^i - \sum_j \gamma^{i,e}_j m^i b_j  - {\mu^{i,e}} \right)^2}
      { \left(\delta_{i,e,{\rm stat}}\,\mu^{i,e}\right)^2+
        \left(\delta_{i,e,{\rm uncor}}\,m^i\right)^2}
 + \sum_j b^2_j\,.
\label{eq:ave}
\end{equation}
Here  ${\mu^{i,e}}$ is the measured value of $\sigma_{\rm red}^{c\bar{c}}(x_i,Q^2_i)$ at an $(x,Q^2)$ point $i$ and $\gamma^{i,e}_j $, $\delta_{i,e,{\rm stat}} $ and 
$\delta_{i,e,{\rm uncor}}$ are the relative
correlated systematic, relative statistical and relative uncorrelated systematic uncertainties,
respectively.
The vector $\pmb{m}$ of quantities $m^i$ expresses the values 
of the combined cross section for each point $i$ and the vector 
$\pmb{b}$ of quantities $b_j$ expresses the shifts 
of the correlated systematic uncertainty sources, $j$, in units of the standard deviation. 
%
%
%
%
Several data sets providing a number of measurements are represented
by a total $\chi^2$ function,
which is built from the sum of the $\chi^2_{{\rm exp},e}$ functions of all data sets
\begin{equation}
\chi^2_{\rm tot} = \sum_e \chi^2_{{\rm exp},e}~. \label{eq:tot}
\end{equation}
The combined reduced cross sections 
are given by the vector $\boldsymbol{m}$ obtained by the minimisation of
$\chi^2_{\rm tot}$
with respect to $\boldsymbol{m}$ and $\boldsymbol{b}$.
With the assumption that the statistical uncertainties are constant 
and that the systematic uncertainties are proportional to 
$m^i$, this minimisation provides an almost unbiased estimator of $\boldsymbol{m}$.

The double differential cross section measurements, used as input for the combination, are available~\cite{webpage} 
with their statistical and  systematic uncertainties. The statistical uncertainties correspond 
to $\delta_{i,e,{\rm stat}}$ in equation~(\ref{eq:ave}). The systematic uncertainties within each measurement 
are classified as either point-to-point correlated or point-to-point uncorrelated, corresponding to 
$\gamma^{i,e}_j$ and $\delta_{i,e,{\rm uncor}}$, respectively. Asymmetric systematic uncertainties 
are symmetrised before performing the combination. The result is found to be insensitive 
to the details of the symmetrisation procedure.

In the present analysis the correlated and uncorrelated systematic uncertainties are predominantly of multiplicative 
nature, i.e. they change proportionally to the central values. In equation~(\ref{eq:ave}) the 
multiplicative nature of these uncertainties is taken into account by multiplying the 
relative errors $\gamma^{i,e}_j$ and $\delta_{i,e,{\rm uncor}}$ by the expectation $m^i$.

In charm analyses the statistical uncertainty is mainly background dominated. 
Therefore it is treated as constant independent of $m^i$. 
To investigate the sensitivity of the result on the treatment of the uncorrelated and, in particular, statistical uncertainty, the analysis is repeated using 
an alternative $\chi^2$ definition in which only correlated 
uncertainties are taken as multiplicative while the uncorrelated uncertainties are treated as constant. 
In a third approach 
the statistical uncertainties are assumed to be proportional to the 
square root of $m^i$. 
The differences between the results obtained from these variations
and the nominal result 
are taken into account as an asymmetric procedural uncertainty 
and are added to the total uncertainty of the combined result in quadrature.

Correlations between systematic uncertainties of different measurements are 
accounted for. Experimental systematic uncertainties are treated as independent between 
H1 and ZEUS. Extrapolation uncertainties due to the variation of the charm quark mass and the 
renormalisation and factorisation scales, charm fragmentation as well as branching fractions 
are treated as correlated. All 
reduced cross section data from H1 and ZEUS are combined in one simultaneous minimisation, 
through which the correlated uncertainties are reduced also at ($Q^2,x$) points where only one measurement exists.

\section{Combined charm cross sections}
\label{sect:results}
The values of the combined cross section \red together with uncorrelated, correlated, 
procedural and total uncertainties are given in table~\ref{tab:result}.
In total, $155$ measurements are combined to $52$ cross-section measurements. 

\begin{figure}[h]
\center

\epsfig{file=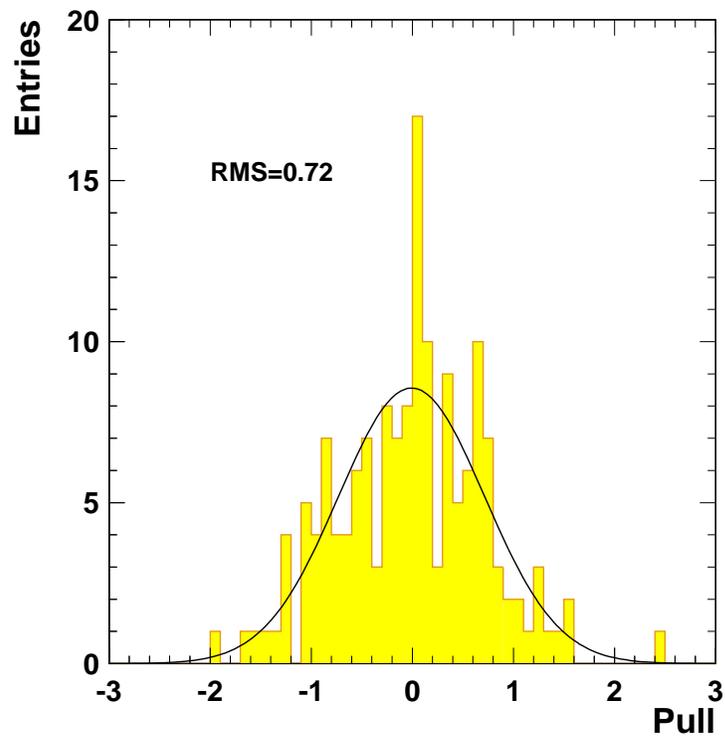,width=0.7\textwidth}
\caption{Pull distribution for the combined data samples (shaded histogram). RMS gives the root mean square of the distribution. The curve shows the result of a binned log-likelihood Gaussian fit.}
\label{fig:pull} 
\end{figure}

The data show good consistency, with a $\chi^2$-value per degree of freedom, $n_{\rm dof}$, of $\chi^2/n_{\rm dof}=62/103$, indicating that the 
uncertainties of the individual measurements have been estimated conservatively. 
The distributions of pulls (as defined in~\cite{DIScomb}) is shown in figure~\ref{fig:pull}.
No significant tensions are observed.
For data with no correlated systematic uncertainties the pulls are expected to follow 
Gaussian distributions with zero mean and unit width. 
Correlated systematic uncertainties lead to narrowed pull distributions. 
%
%

There are in total $48$ sources of correlated systematic uncertainty, including global normalisations, characterising the 
separate data sets. The shifts and the reduction of the correlated uncertainties are given in table~\ref{tab:systematic_shifts}.
None of these systematic sources shifts by more than $1.2 \,\sigma$ of the nominal value in the averaging procedure. 
The influence of several correlated systematic uncertainties is reduced significantly in the result.
For example the uncertainties from the vertex analyses due to the light quark background (H1) and due to the tracking (ZEUS) are reduced by almost a factor of two. 
The reductions can be traced mainly to the different charm tagging methods, and to the requirement that 
different measurements probe the same cross section at each ($x,Q^2$) point. 
In addition, for certain kinematic regions one measurement has superior precision and the less precise ones follow its trend through the fit. 
The reduction of systematic uncertainties propagates to the other average 
points, including those which are based solely on the less precise measurements.

\begin{figure}[h]
\center
\epsfig{file=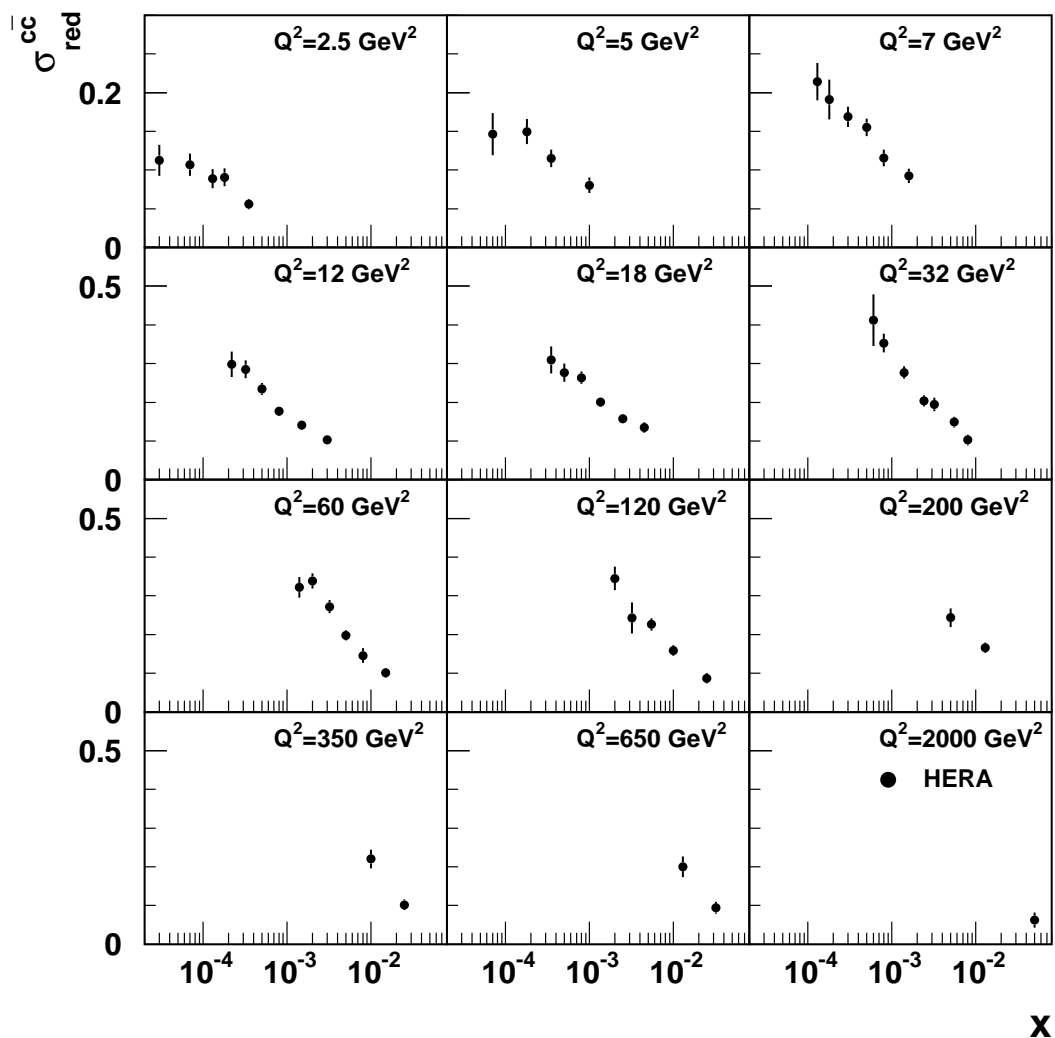,width=1.0\textwidth}
\caption{Combined  reduced cross sections \red as a function of $x$ for fixed values of $Q^2$. The error bars represent the total uncertainty including uncorrelated, correlated and procedural uncertainties added in quadrature.}
\label{fig:combined} 
\end{figure}

\begin{figure}[h]
\center
\epsfig{file=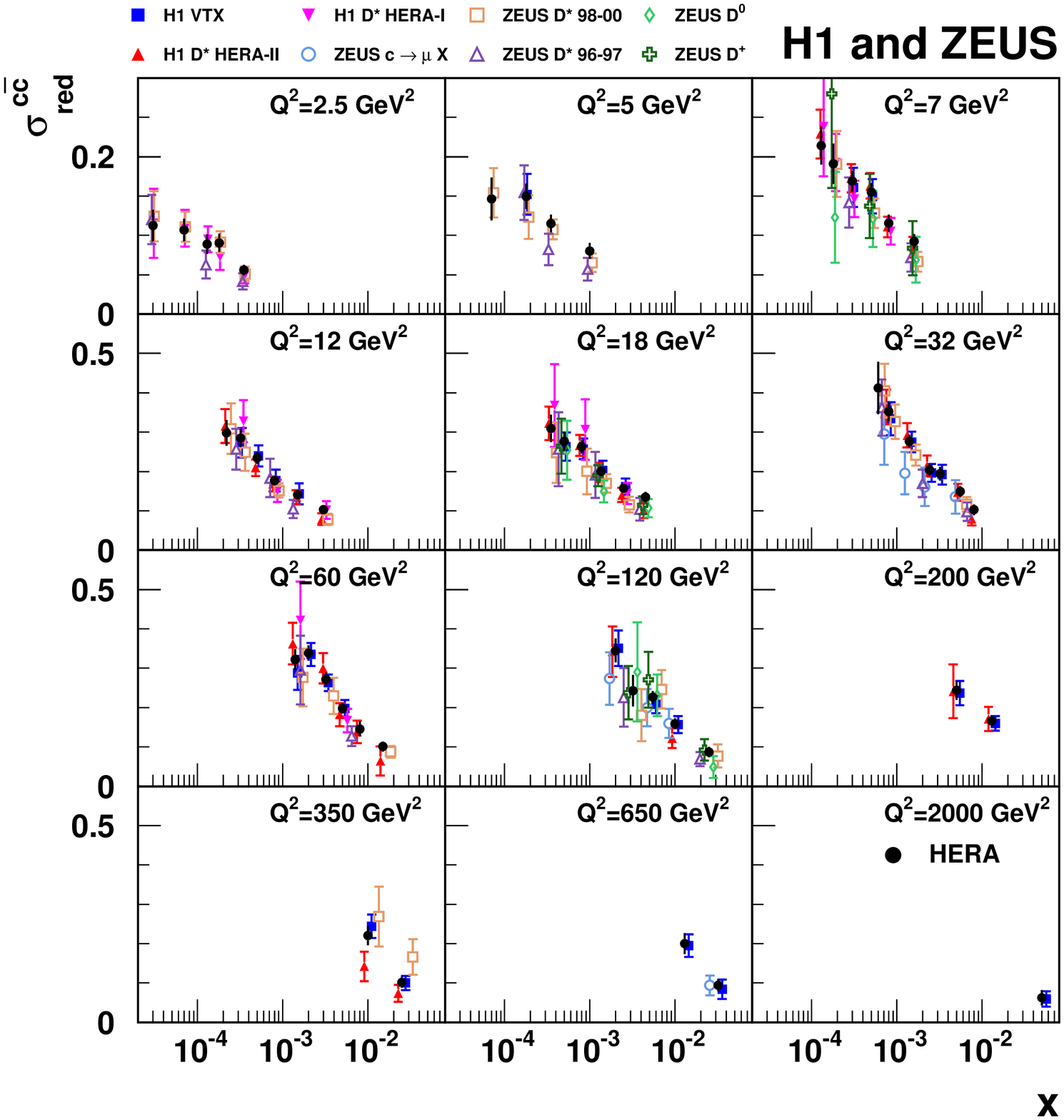,width=1.0\textwidth}
\caption{Combined reduced cross sections \red (filled circles) as a function of $x$ for fixed values of $Q^2$. The error bars represent the total uncertainty including uncorrelated, correlated and procedural uncertainties added in quadrature. For comparison, the input data are shown: the H1 measurement based on lifetime information of inclusive track production is represented by closed squares; the H1 measurements based on reconstruction of $D^*$ mesons in HERA-I / HERA-II running periods are denoted by filled up (down) triangles; the ZEUS measurement using semileptonic decays into muons is represented by open circles; the ZEUS measurements based on reconstruction of $D^*$ mesons are depicted by open squares (open triangles) for data collected in 1998-2000 (1996-1997) years; the ZEUS measurements based on reconstruction of $D^0$ ($D^+$) mesons are shown by open diamonds (crosses). For presentation purpose each individual measurement is shifted in $x$.}
\label{fig:combined_vs_input} 
\end{figure}

\begin{figure}[h]
\center
\epsfig{file=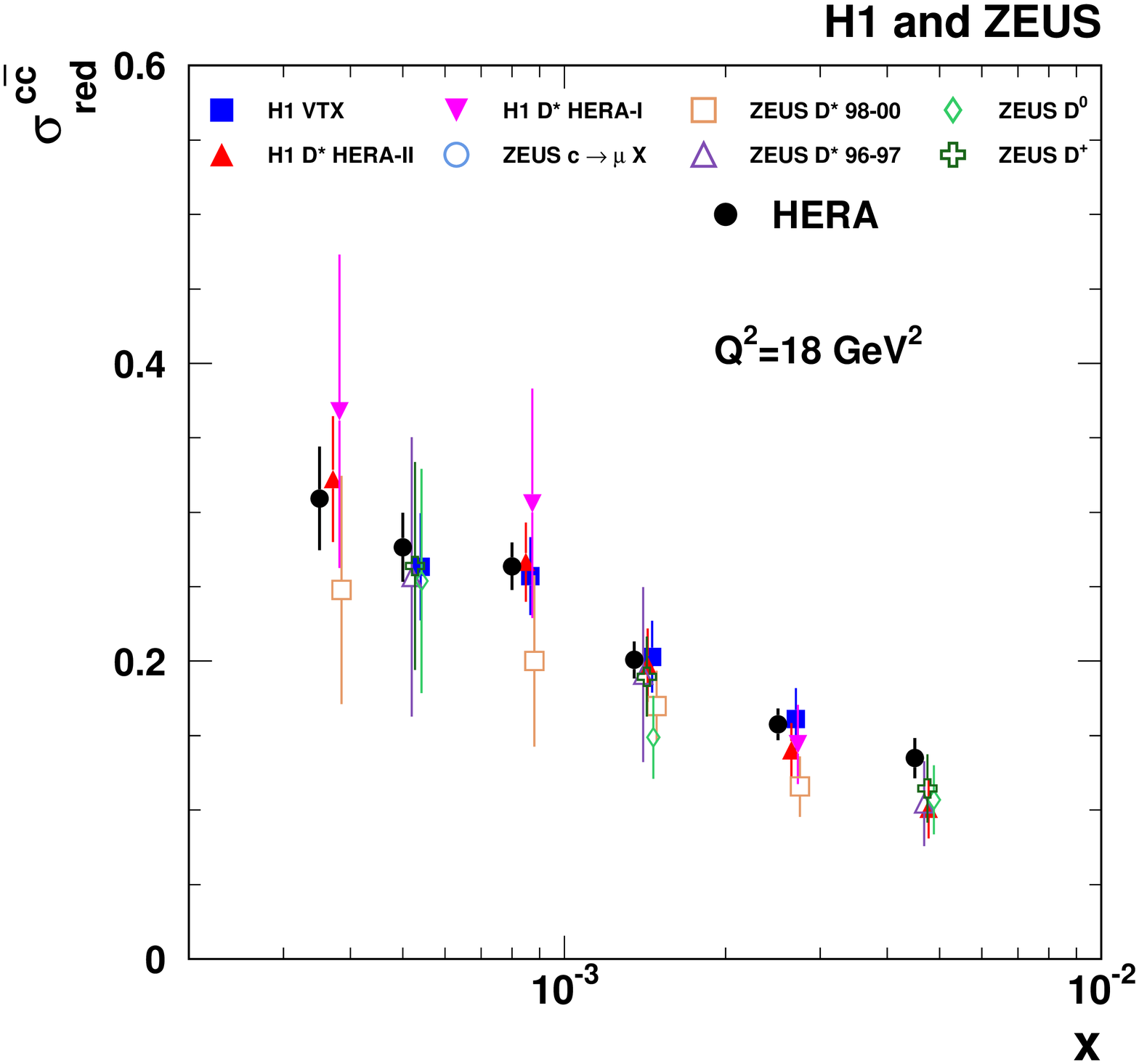,width=1.0\textwidth}
\caption{Combined reduced cross sections \red (filled circles) as a function of $x$ for $Q^2=18$ GeV$^2$. The error bars represent the total uncertainty including uncorrelated, correlated and procedural uncertainties added in quadrature. For comparison, the input data are shown. For further details see figure \ref{fig:combined_vs_input}.}
\label{fig:combined_vs_input_zoom} 
\end{figure}

The cross section tables of the input data sets used in the analysis (see section \ref{Sect:combination}) together with the full information of the correlations among these cross section measurements can be found elsewhere~\cite{webpage}. 
The combined reduced cross section is presented in figure~\ref{fig:combined} as a function of $x$, in bins of $Q^2$, and compared to the input H1 and ZEUS data 
in figure~\ref{fig:combined_vs_input}. The combined data are significantly more precise than any of the individual 
input data sets. This is illustrated in figure~\ref{fig:combined_vs_input_zoom}, where 
the measurements for $Q^2=18$~GeV$^2$ are shown. 
The uncertainty of the combined results is $10\%$ on average and reaches 
$6\%$ in 
the region of small $x$ and medium $Q^2$. This is an improvement of about a factor of 2 with respect to each of the 
most precise data sets in the combination.
\section{Comparison to theoretical predictions}
\label{sect:comparisons}

Before proceeding to the QCD analysis including these 
data, it is instructive to compare them to various QCD predictions produced by 
different theory groups, for which the parameters are listed in 
table \ref{tab:theories}. 
This comparison tests the interplay between the 
gluon and/or heavy flavour PDFs as obtained in different schemes and the 
charm treatment within each scheme (section 2), 
as well as the related choice of the 
central value for the respective charm mass parameter. 
Some of the findings in this section can be cross-related to corresponding
more detailed NLO QCD studies in section \ref{sect:qcdana}.    
In addition, the effect of NNLO corrections is studied here. 
The full calculation of the 
heavy quark coefficient functions is available at $O(\alpha_s^2)$ only. 
The $O(\alpha_s^3)$ corrections listed in table \ref{tab:theories} correspond 
to partial resummation corrections applied in some 
kinematic ranges of charm production. 
Most predictions already contain some measured charm data from previous 
publications as input (see table \ref{tab:theories} for details).

\begin{sidewaysfigure}[h]
\center
\setlength{\unitlength}{1cm}
\begin{picture}(24,12.)
\put(0.,0.){\epsfig{file=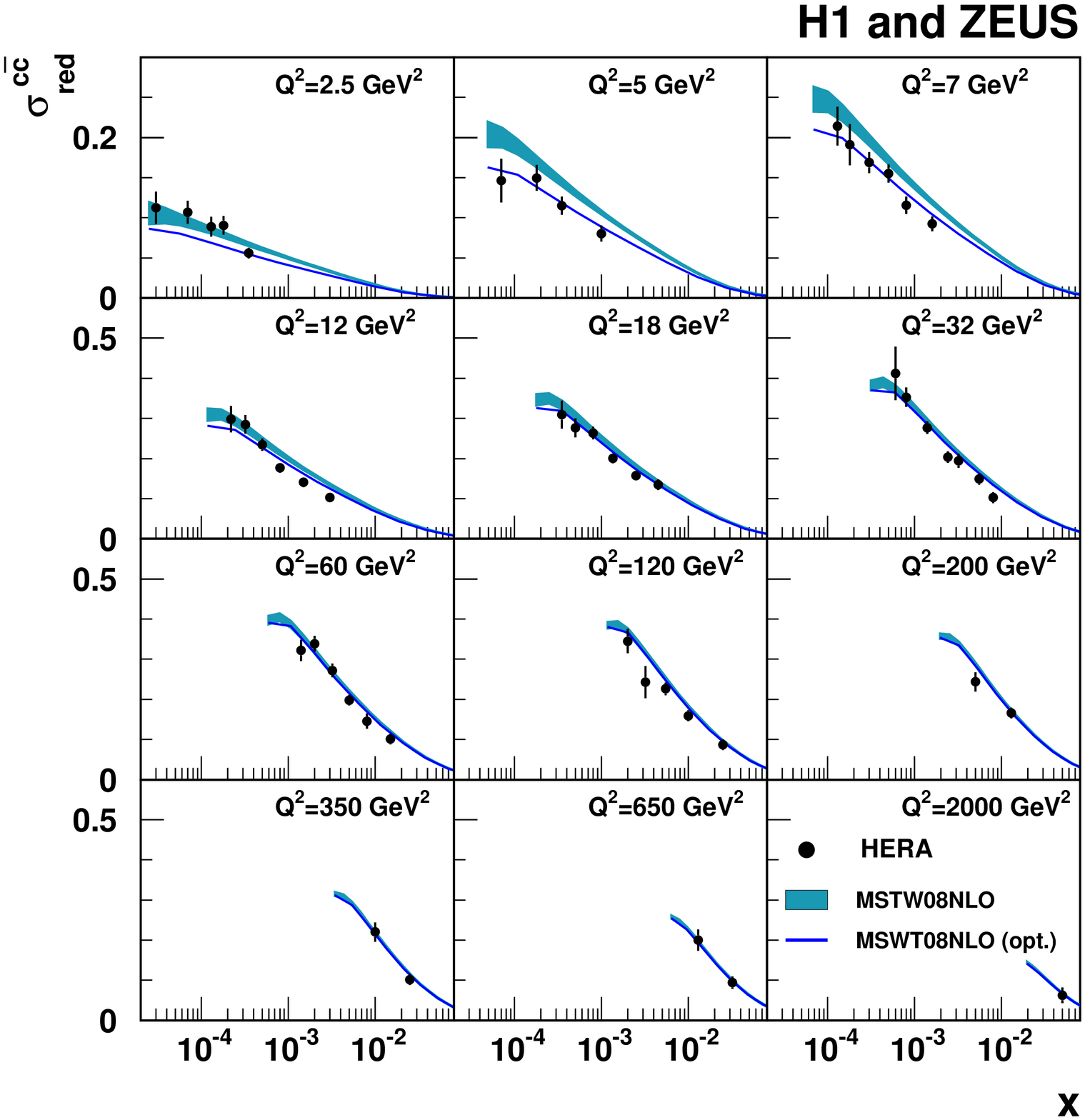,width=0.52\textwidth}}
\put(12.,0.){\epsfig{file=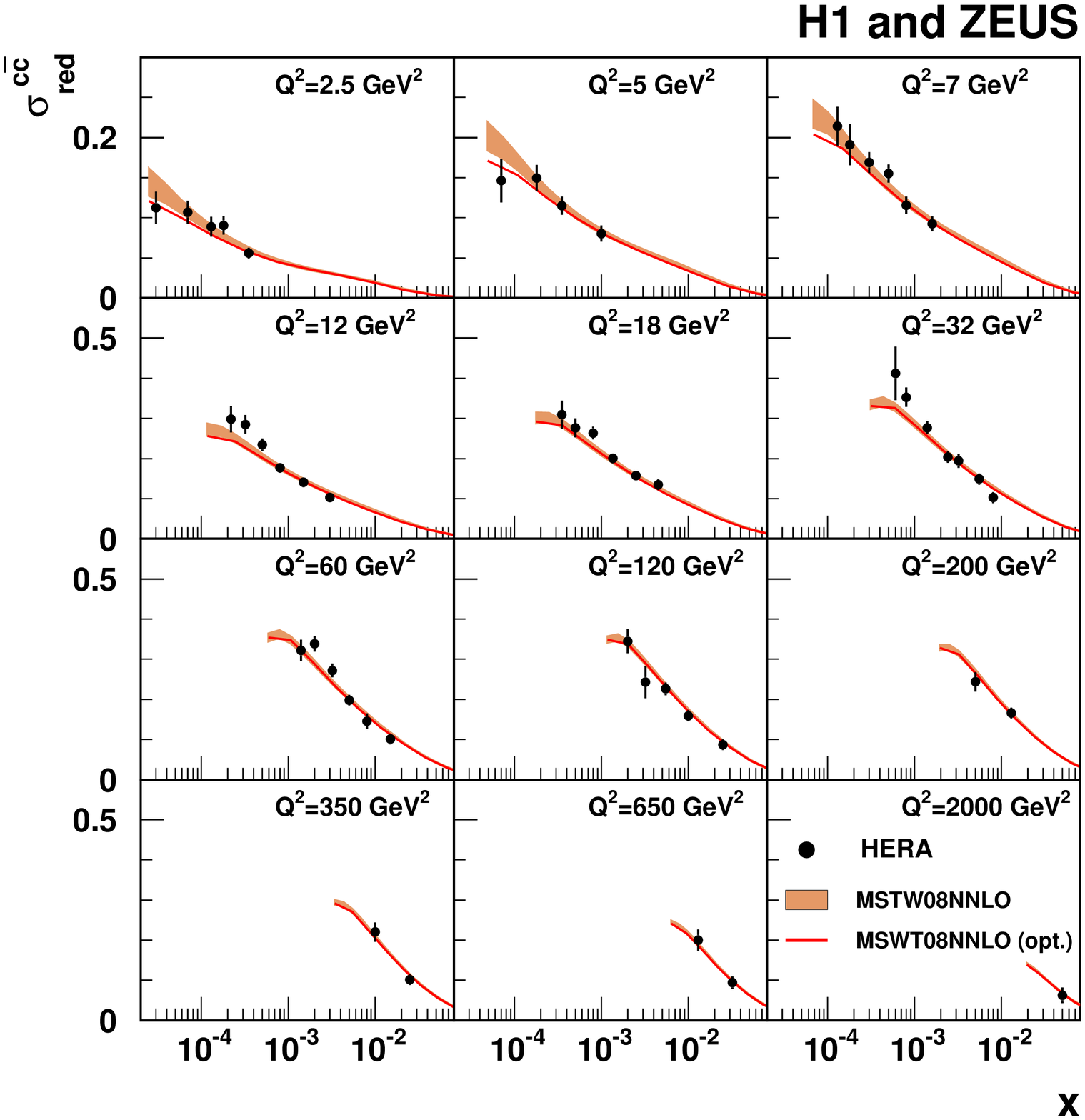,width=0.52\textwidth}}
\end{picture}
\setlength{\unitlength}{1cm}
\caption{Combined reduced cross sections \red (filled circles) as a function of $x$ for fixed values of $Q^2$. The error bars represent the total uncertainty including uncorrelated, correlated and procedural uncertainties added in quadrature. The data are compared to MSTW predictions at NLO (left panel) and NNLO (right panel). The predictions obtained using the standard (optimised) parametrisation are represented by 
the shaded bands (solid lines). The uncertainties for the optimised parametrisation are not yet available.}
\label{fig:combined_mstw} 
\end{sidewaysfigure}

In figure~\ref{fig:combined_mstw} the reduced cross section \red 
is compared with 
predictions of the MSTW group in the GM-VFNS at NLO and NNLO, 
using the RT standard~\cite{rt_std} and the RT optimised~\cite{rt_opt} interpolation procedure 
of the cross section at the charm production threshold.
At NLO, the optimised prediction tends to describe the data better than 
the standard one at lower $Q^2$.
The description of the data is improved in NNLO compared to NLO. 

\begin{figure}[h]
\center
\epsfig{file=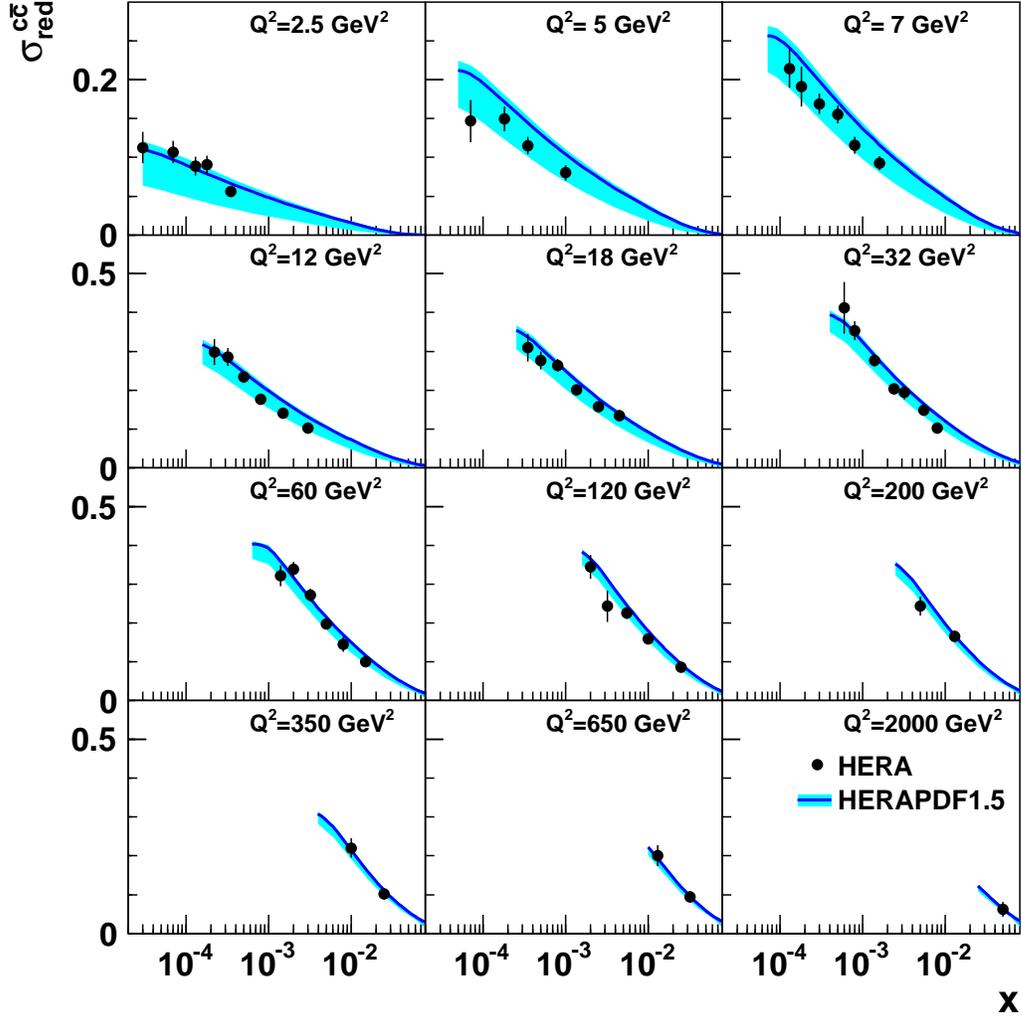,width=1.\textwidth}
\setlength{\unitlength}{1cm}
\caption{Combined reduced cross sections \red (filled circles) as a function of $x$ for fixed values of $Q^2$. The error bars represent the total uncertainty including uncorrelated, correlated and procedural uncertainties added in quadrature. The data are compared to the NLO predictions based on HERAPDF1.5 extracted in the RT standard scheme. The line represents the prediction using $\mct =1.4$~GeV. The uncertainty band shows the full PDF uncertainty which is dominated by the variation of \mct.}
\label{fig:combined_HERAPDF1.5} 
\end{figure}

In figure~\ref{fig:combined_HERAPDF1.5} the data are compared to the NLO predictions based on HERAPDF1.5 \cite{ref:HERAPDF1.5} extracted in the RT standard scheme  
using as inputs the published HERA-I \cite{DIScomb} and the preliminary HERA-II combined inclusive DIS data. For the central PDF set a charm quark mass parameter $\mct=1.4$~GeV is used. 
The uncertainty bands of the predictions reflect the full uncertainties on the HERAPDF1.5 set. 
They are dominated by the uncertainty on \mct~which is varied between $1.35$~GeV and $1.65$~GeV \cite{DIScomb}. Within these uncertainties the HERAPDF1.5 predictions describe the data well.
The central predictions are very similar to those of the MSTW group for the same scheme. 

\begin{sidewaysfigure}[h]
\center
\setlength{\unitlength}{1cm}
\begin{picture}(24,12.)
\put(0.,0.){\epsfig{file=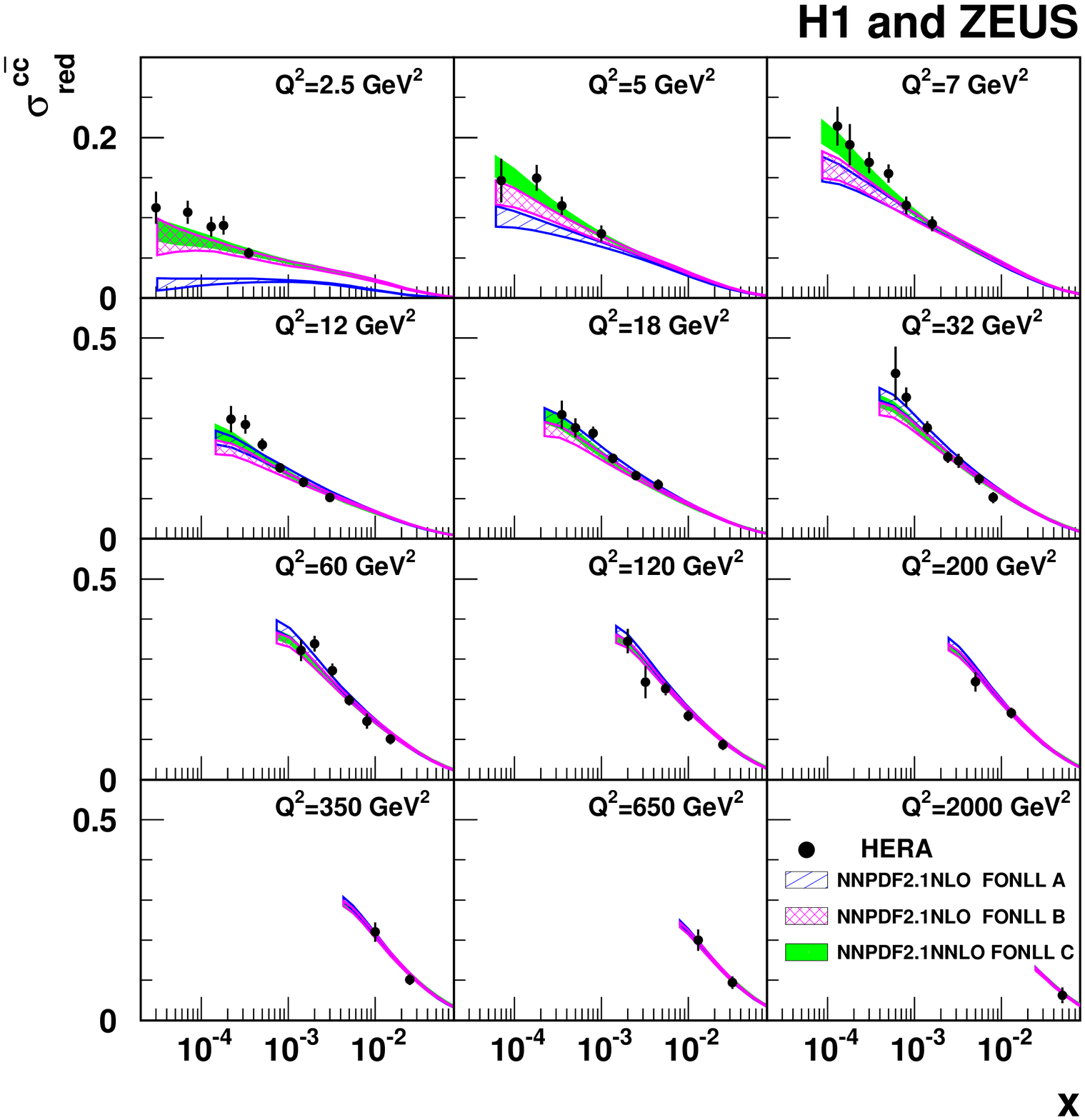,width=0.52\textwidth}}
\put(12.,0.){\epsfig{file=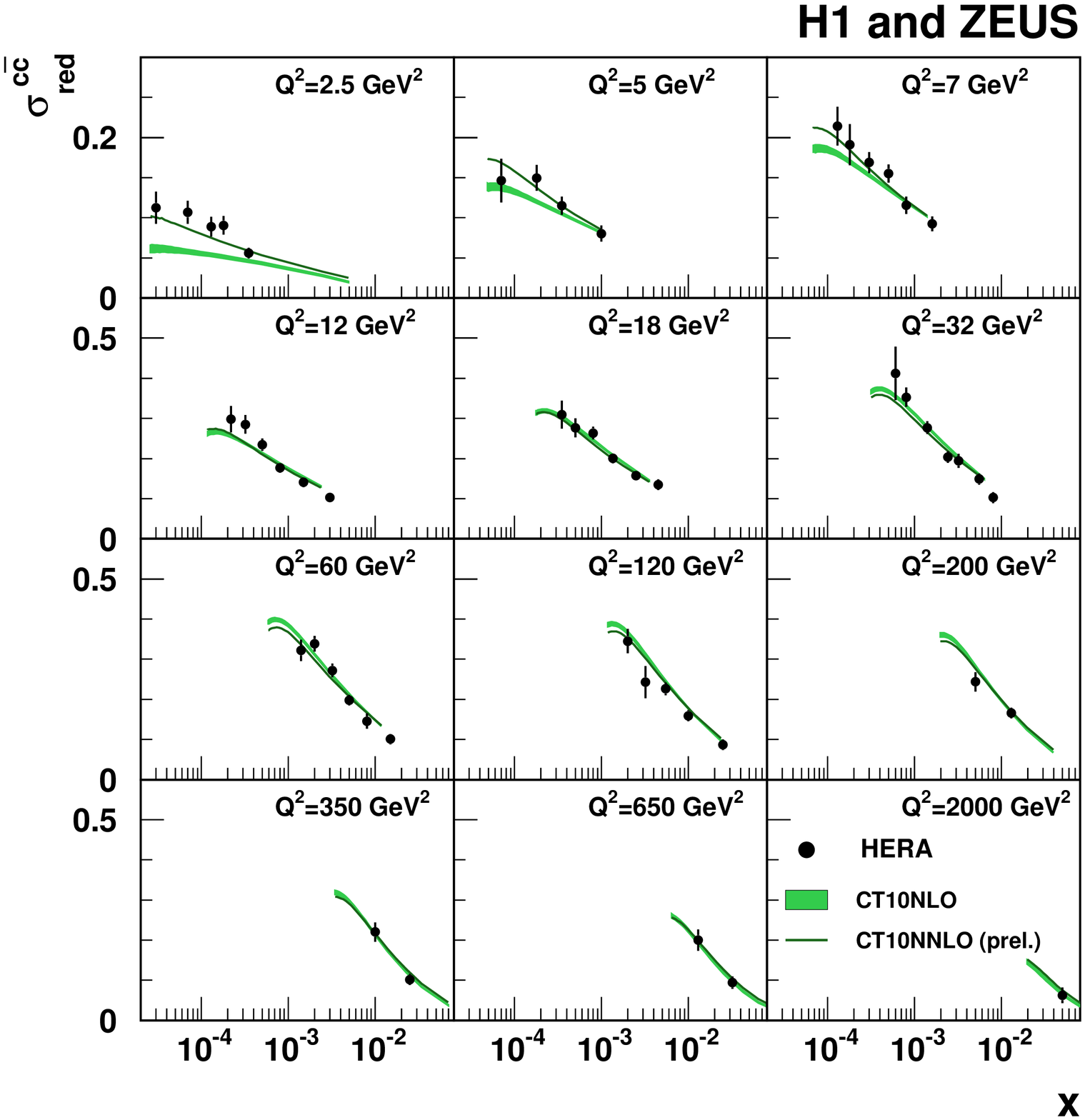,width=0.52\textwidth}}
\end{picture}
\setlength{\unitlength}{1cm}
\caption{Combined reduced cross sections \red (filled circles) as a function of $x$ for fixed values of $Q^2$. The error bars represent the total uncertainty including uncorrelated, correlated and procedural uncertainties added in quadrature. The data are compared to predictions by the NNPDF group (left panel) and the CTEQ group (right panel). The predictions from NNPDF2.1 in FONNL-A, -B and -C schemes are shown with their uncertainties (bands with different hatch styles). The CT10 NLO prediction with its uncertainties is shown by the shaded bands. The uncertainties on the CT10 NNLO (prel.) predictions are not yet available.}
\label{fig:combined_nnpdf_ct} 
\end{sidewaysfigure}

In figure~\ref{fig:combined_nnpdf_ct} the data are compared to the 
predictions in the GM-VFNS by the NNPDF and CT collaborations. 
Both the NNPDF FONLL-A~\cite{FONLL} and  FONLL-B~\cite{fonllb_and_c} 
predictions describe 
the data fairly well at higher $Q^2$, while they fail to describe the 
data at lower $Q^2$. The description of the data at lower $Q^2$ is improved 
in the FONLL-C~\cite{fonllb_and_c} scheme.
The CT predictions~\cite{ct10f3,ct12nnlo} are based on the S-ACOT-$\chi$ heavy 
quark scheme. 
The NLO prediction, which 
is very similar to 
the FONLL-A scheme, describes the data well for 
$Q^2>5$ GeV$^2$ but fails 
to describe the data at lower $Q^2$. 
Similar to the FONNL-C case 
the description of the data 
improves significantly at NNLO.

\begin{figure}[h]
\center
\epsfig{file=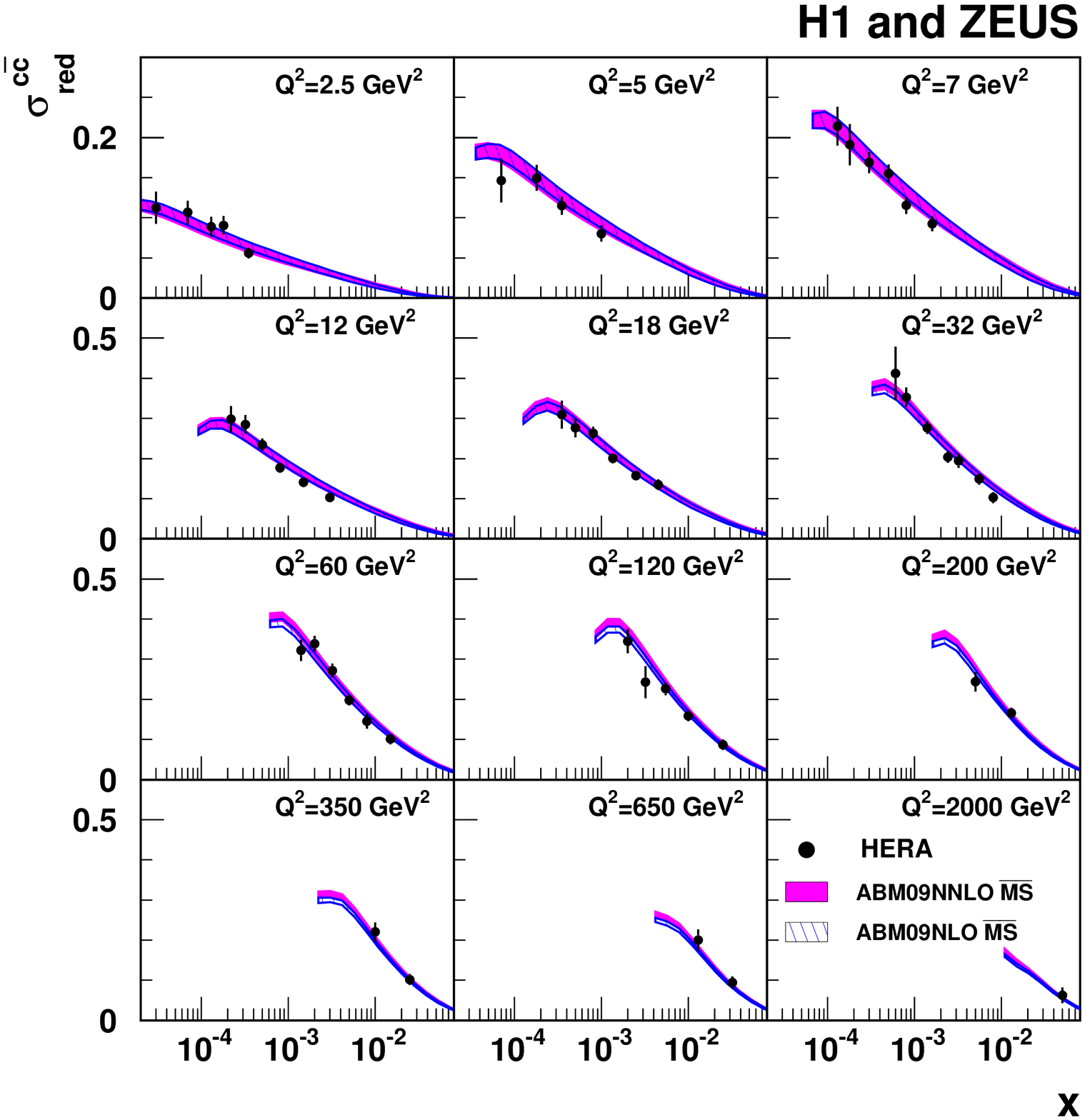,width=01.\textwidth}
\setlength{\unitlength}{1cm}
\caption{Combined reduced cross sections \red (filled circles)  as a function of $x$ 
for fixed values of $Q^2$. The error bars represent the total uncertainty including uncorrelated, correlated 
and procedural uncertainties added in quadrature. The data are compared to predictions of the ABM group at 
NLO (hashed band) and NNLO (shaded band) in FFNS using the $\overline{\text{MS}}$ definition for the charm quark mass.}
\label{fig:combined_abkm} 
\end{figure}

In figure~\ref{fig:combined_abkm} the data are compared to the prediction
of the ABM group in FFNS at NLO and NNLO, based on the running-mass scheme for 
both the coefficient functions and the PDFs ~\cite{ABKMMSbar,abkm}. 
The uncertainties on the prediction include 
the uncertainties on $m_c$, which dominate at small $Q^2$. 
The predictions at NLO and NNLO are very similar and describe the data well in the whole kinematic range of 
the measurement. 

In summary, the best description of the data is achieved by predictions 
including partial $O(\alpha_s^3)$ corrections (MSTW NNLO and ABM NNLO).
The predictions including $O(\alpha_s^2)$ terms in all
parts of the calculation (NNPDF FONLL C, CT NNLO and ABM NLO) as well as the 
MSTW NLO optimal scheme also agree well with the data. 
The largest deviations are observed for predictions based on $O(\alpha_s)$ 
terms only (NNPDF FONLL A and CT NLO).
As investigated in the next section,
further differences can be partially explained 
by the different choices for the value of the respective charm quark mass parameter \mct.

\section{QCD analysis}
\label{sect:qcdana}

The combined H1 and ZEUS inclusive $ep$ neutral current and 
charged current DIS cross sections have been used previously 
to determine the HERAPDF1.0 parton density functions. 
In the current paper a combined NLO QCD analysis is performed using the reduced charm cross section together with 
the combined inclusive DIS cross sections \cite{DIScomb}.
Since the charm contribution to the inclusive DIS cross section is sizeable and reaches up to $\approx 30\%$ at high $Q^2$, this combined analysis is expected to reduce the uncertainties related to charm production inherent in all PDF extractions. 
In particular, the r\^ole of the charm quark mass $m_c(m_c)$ or the charm quark mass parameter $M_c$, depending on the heavy flavour scheme, is investigated within all schemes discussed in section~\ref{sect:theory}.

The analysis is performed with the HERAFITTER~\cite{herafitter} program, which is based on the NLO DGLAP 
evolution scheme~\cite{dglap} as implemented in QCDNUM~\cite{qcdnum}.
The invariant mass of the hadronic system is restricted to $W>15$~GeV, and the 
Bjorken scaling variable $x$ is limited by the data to $x \le 0.65$. In this kinematic range target mass corrections 
and higher twist contributions are expected to be small. In addition,
the analysis is restricted to data with 
$Q^2>Q^2_{min}=3.5$~GeV$^2$ to assure the applicability of pQCD. The consistency of the input data sets and the good
control of the systematic uncertainties enable the determination of the experimental uncertainties on 
the PDFs using the $\chi^2$ tolerance of $\Delta\chi^2=1$.  

The following independent PDFs are chosen in the fit procedure:
$xu_v(x)$, $xd_v(x)$, $xg(x)$ and 
$x\overline{U}(x)$, $x\overline{D}(x)$, where $x\overline{U}(x) = x\overline{u}(x)$, and
$x\overline{D}(x) = x\overline{d}(x) +x\overline{s}(x)$.
Compared to the HERAPDF1.0 analysis, a more flexible parameterisation with $13$ free parameters is used.
At the starting scale $Q_0$ of the QCD evolution, the PDFs are parametrised as follows:
\begin{eqnarray}
xg(x)&=&A_g x^{B_g}\cdot (1-x)^{C_g}-A'_g x^{B'_g}\cdot (1-x)^{C'_g},\\
xu_v(x)&=& A_{u_v} x^{B_{u_v}}\cdot (1-x)^{C_{u_v}}\cdot (1+E_{u_v}x^2)\label{puv},\\
xd_v(x)&=& A_{d_v} x^{B_{d_v}}\cdot (1-x)^{C_{d_v}}\label{pdv},\\
x\overline{U}(x)&=& A_{\overline{U}} x^{B_{\overline{U}}}\cdot (1-x)^{C_{\overline{U}}},\\
x\overline{D}(x)&=& A_{\overline{D}} x^{B_{\overline{D}}}\cdot (1-x)^{C_{\overline{D}}}\label{pD}.
\end{eqnarray}
The normalisation parameters $A_g$, $A_{u_v}$, $A_{d_v}$ are constrained by the sum rules. 
The parameter $B_{\overline{U}}$ is set to $B_{\overline{D}}$ 
and the constraint $A_{\overline{U}}=A_{\overline{D}}(1-f_s)$, with $f_s$ being the strangeness fraction at the starting scale, ensures the same normalisation for the $\overline{u}$ and $\overline{d}$ densities for $x\rightarrow 0$. The strangeness fraction is set to $f_s=0.31$, as obtained from neutrino-induced di-muon production \cite{fsnu}. 
To ensure a positive gluon density at large $x$, the parameter $C'_g$ is set to $25$, in accordance with 
~\cite{rt_std}.

The study involves variations of the charm mass parameter down to 
$\mct=1.2$~GeV with the exception of the S-ACOT-$\chi$ scheme for which the \mct~scan starts at $\mct=1.01$~GeV. Since the starting scale $Q_0$ has to be smaller than $M_c$, the
 fits are performed with setting $Q^2_0=1.4$~GeV$^2$ and $Q^2_0=1.0$~GeV$^2$, respectively. 
In order to keep the variation of $M_c$ independent from a $Q_0$ variation,
this value for 
$Q_0$ is chosen irrespectively of the actual value of $M_c$ used during the variation procedure. 

The renormalisation and factorisation scales are set to $Q$ for the VFNS and for the light quark contribution in the FFNS and to $\sqrt{Q^2+4m_{c,b}^2}$ for the contribution of a heavy quark in the FFNS. 
For the strong coupling constant the values $\alpha_s(M_Z) =  0.1176$~\cite{pdg} and $\alpha_s^{n_f=3}(M_Z)=0.105$ with $n_f=3$ active flavours in the proton are used for the VFNS and for the FFNS, respectively.
The definition of $F_{2,L}$ and the $\alpha_s$ order of the calculation are 
the same as those listed for the respective scheme in 
table \ref{tab:theories} at NLO (ACOT-full and ZM-VFNS see S-ACOT-$\chi$).

\begin{figure}[]
\center
\epsfig{file=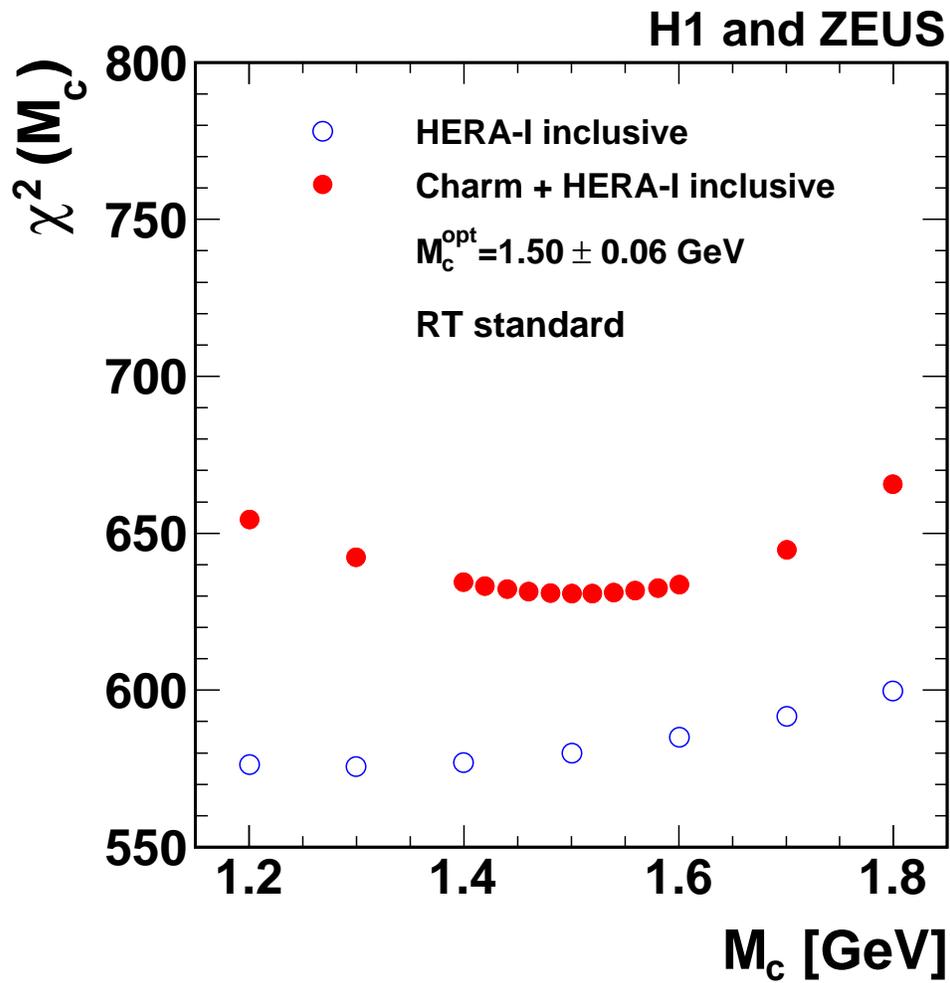,width=1.\textwidth}
\caption{The values of $\chi^2(M_c)$ for the PDF fit to the combined HERA DIS data in the RT standard scheme. 
The open symbols indicate the results of the fit to inclusive 
DIS data only. The results of the fit including the combined charm data are shown by 
filled symbols.}
\label{fig:rtstd} 
\end{figure}

For each heavy flavour scheme a number of PDF fits is performed with varying \mct~from 1.2 GeV to 1.8 GeV. For each fit 
the $\chi^2(\mct)$ value is calculated and the optimal value, \mcto, of the charm quark mass parameter in a given scheme is subsequently determined from a parabolic fit 
of the form
\begin{equation}
  \chi^2(\mct) = \chi^2_{\rm min} + \left(\frac{\mct - \mcto}{\sigma(\mcto)}\right)^2\,
\label{mcfit}
\end{equation}
to the $\chi^2(\mct)$ values.
Here $\chi^2_{\rm min}$ is the $\chi^2$ value at the minimum and $\sigma(\mcto)$ is the fitted experimental uncertainty on $\mcto$.
The procedure of this $\chi^2$-scan is illustrated in figure~\ref{fig:rtstd} for the standard RT scheme when fitting only the inclusive HERA-I DIS data and when fitting these data together with \red. 
The inclusive NC and CC cross sections from HERA-I alone only weakly constrain $\mct$; the value of $\chi^2(\mct)$ varies only slowly with $\mct$. 
Once the charm data are included, a clear minimum is observed, which then determines \mcto.
 
The systematic uncertainties on 
\mcto~are calculated from 
the following variations of the model assumptions:
\begin{itemize}

\item{\bf the strangeness fraction} is varied in the range $0.23<f_s<0.38$. In a recent publication the ATLAS collaboration \cite{fsatlas} has observed $f_s=0.5$. This value of $f_s$ is also tested and found to have only a negligible effect on the determination of \mcto.

\item{\bf the \boldmath $b$-quark mass} is varied by $\pm 0.25$~GeV around the central value of $4.75$~GeV.
 
\item {\bf the minimum \boldmath $Q^2$ value for data used in the fit, {\boldmath $Q^2_{\rm min}$}}, is varied for the inclusive data from $Q^2_{\rm min}=3.5$~GeV$^2$ to  $5.0$~GeV$^2$. 
For the charm data this variation is not applied because it would significantly reduce the sensitivity of the analysis on \mct. However,
the full difference on the fitted value \mcto~obtained by using the cuts
$Q^2_{\rm min}=3.5$~GeV$^2$ or $Q^2_{\rm min}=5$~GeV$^2$, 
is then taken as symmetric uncertainty due to the variation of $Q_{\rm min}^2$.

\item{\bf the parameterisation uncertainty} is estimated similarly to the HERAPDF1.0 procedure.
To all quark density functions an additional parameter is added one-by-one such that the parameterisations are changed in equation (\ref{puv}) from $A\cdot x^B\cdot(1-x)^C\cdot(1+Ex^2)$ to $A\cdot x^B\cdot(1-x)^C\cdot(1+Dx+Ex^2)$ and in equations (\ref{pdv}-\ref{pD}) from $A\cdot x^B\cdot(1-x)^C$ to either $A\cdot x^B\cdot(1-x)^C\cdot(1+Dx)$ or $A\cdot x^B\cdot(1-x)^C\cdot(1+Ex^2)$.
Furthermore, the starting scale $Q_0$ is varied to $Q_0^2=1.9$~GeV$^2$. The full difference on the fitted value \mcto, obtained by using $Q_0^2=1.9$~GeV$^2$ and $Q_0^2=1.4$~GeV$^2$ is then taken as symmetric uncertainty due to the variation of the starting scale $Q_0$. 
The total parameterisation uncertainty is obtained taking the largest difference  in \mcto~of the above variations with respect to \mcto~for the standard parameterisation.

\item{\bf the strong coupling constant {\boldmath $\alpha_s (M_Z)$}} is varied by $\pm 0.002$. 
\end{itemize}

For each scheme the assumptions in the fits are varied one by one and the corresponding $\chi^2$ scan as a function of \mct~is performed.
The difference between \mcto\ obtained for the default assumptions and the result of each variation is taken as the corresponding uncertainty.
The dominant contribution 
arises from the variation of $Q^2_{\rm min}$, while the remaining model and parameterisation uncertainties are small 
compared to the experimental error.

\subsection{Extraction of {\boldmath \mcto} in the VFNS}
\label{mcvfns}
The following implementations of the GM-VFNS are considered: ACOT full~\cite{acot} as used for the 
CTEQHQ releases of PDFs; S-ACOT-$\chi$~\cite{s_acot_chi} as used for the latest CTEQ releases of PDFs, and 
for the FONLL-A scheme \cite{FONLL} used by NNPDF;  
the RT standard scheme~\cite{rt_std} as used for the MRST and MSTW releases of PDFs, as well as the RT optimised 
scheme providing a smoother behaviour across thresholds~\cite{rt_opt}. The ZM-VFNS as implemented by 
the CTEQ group~\cite{acot} is also used for comparison. In all schemes, the onset of the heavy quark PDFs is controlled by 
the parameter \mct\, in addition to the kinematic constraints. 

\begin{figure}[]
\center
\epsfig{file=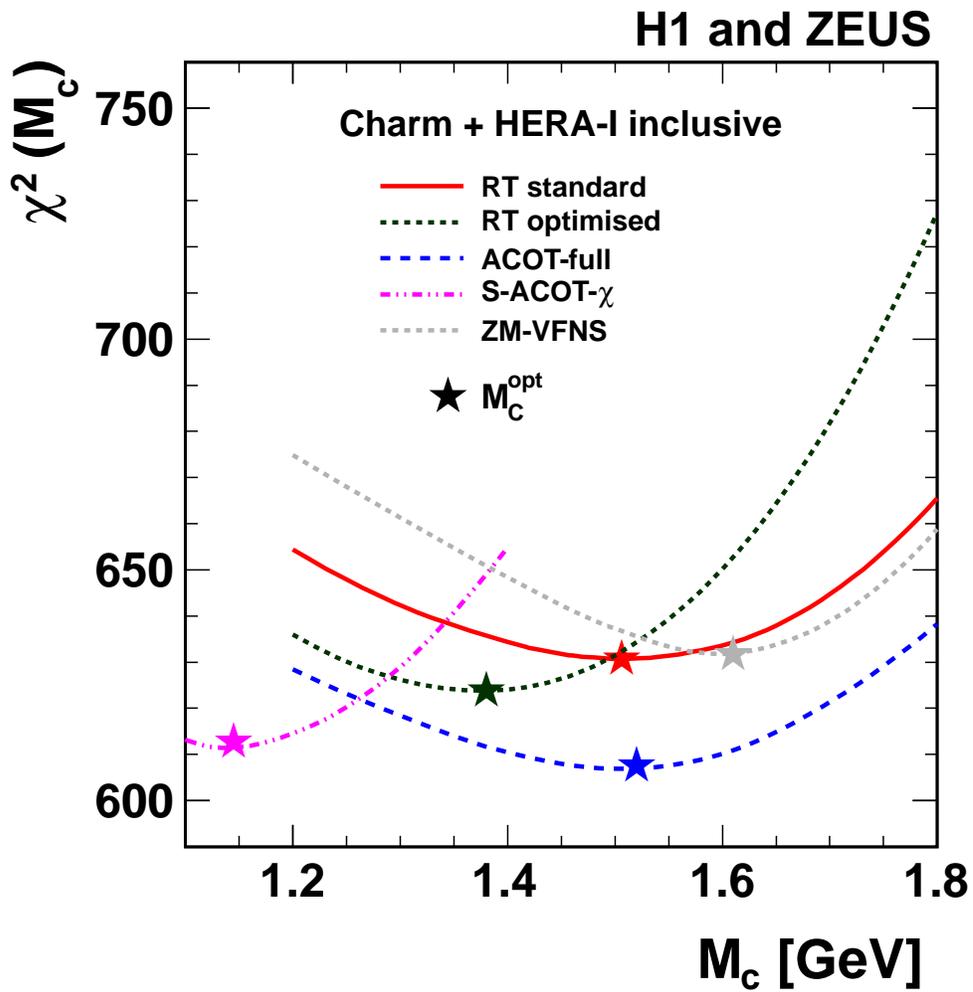,width=1.\textwidth}
\caption{The values of $\chi^2(M_c)$ for the PDF fit to the combined HERA inclusive DIS and charm measurements. 
 Different heavy flavour schemes are used in the fit 
and presented by lines with different styles. The values of $\mcto$ for each scheme are indicated by the stars.}
\label{fig:scharmscan_allschemes} 
\end{figure}

In figure~\ref{fig:scharmscan_allschemes} the $\chi^2$-values as a function of \mct~obtained from PDF fits to the inclusive HERA-I data and the combined charm data are shown for all schemes considered. Similar minimal $\chi^2$-values are observed for the different schemes, albeit at quite different values of \mcto.
In table~\ref{tab:mco}
the resulting values of \mcto~are given together with the uncertainties, the corresponding total $\chi^2$ and the $\chi^2$-contribution from the reduced charm cross section measurements. The ACOT-full scheme yields the best global $\chi^2$, while the best partial $\chi^2$ for the
charm data is obtained using the RT optimised scheme. The fits in the S-ACOT-$\chi$ scheme result in a very low value of \mcto~ as compared to the other schemes.  

\begin{figure}[hhh]
\center
\epsfig{file=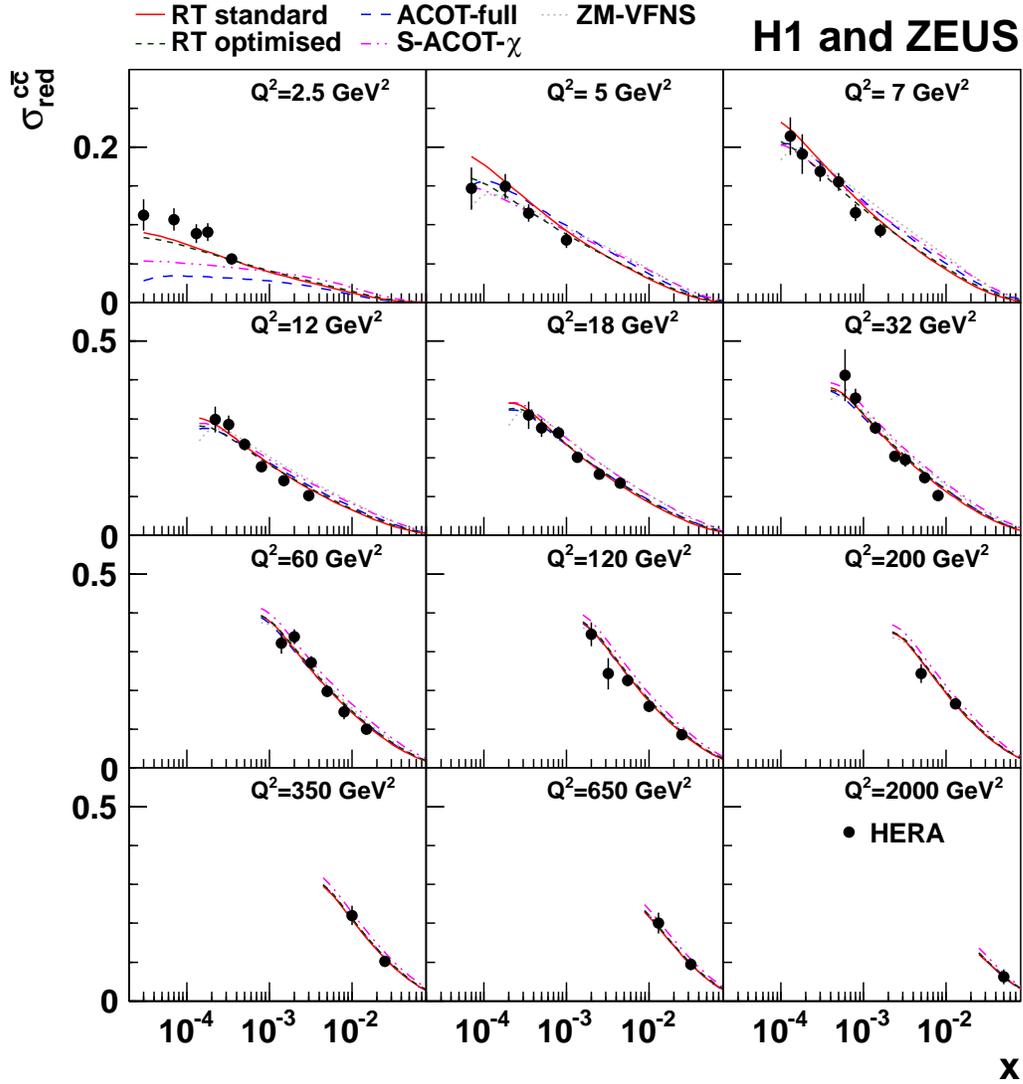,width=1.\textwidth}
\caption{Combined measurements of \red as a function of $x$ for given values of $Q^2$ is shown by filled symbols. The error bars represent the total uncertainty including uncorrelated, correlated and procedural uncertainties added in quadrature. The data are compared to the results of the fit using different variants of the VFNS (represented by lines of different styles) choosing $M_C=\mcto$. The cross section prediction of the ZM-VFNS vanishes for $Q^{2}=2.5~GeV^2$. }
\label{fig:data5} 
\end{figure}

In figure~\ref{fig:data5} the NLO VFNS predictions for \red based on the PDFs evaluated using \mct$=$\mcto\  
of the corresponding scheme are compared to the data. 
In general the data are better described than when using the default values for \mct~and the predictions of the different schemes become very similar for $Q^2\ge 5$~GeV$^2$. 
Even the ZM-VFNS, which includes mass effects only indirectly \cite{acot}, yields an equally good description of the \red~as the GM-VFNS, although it 
fails to describe more differential distributions of $D^{*\pm}$ meson production and the lowest $Q^2$ bin in figure~\ref{fig:data5}, for which the ZM-VFNS cross section prediction is zero.

\subsection{Impact of the charm data on PDFs}

\begin{sidewaysfigure}[hhh]
\center
\unitlength1cm
\begin{picture}(24.,12.)
\put(0.,0.){\epsfig{file=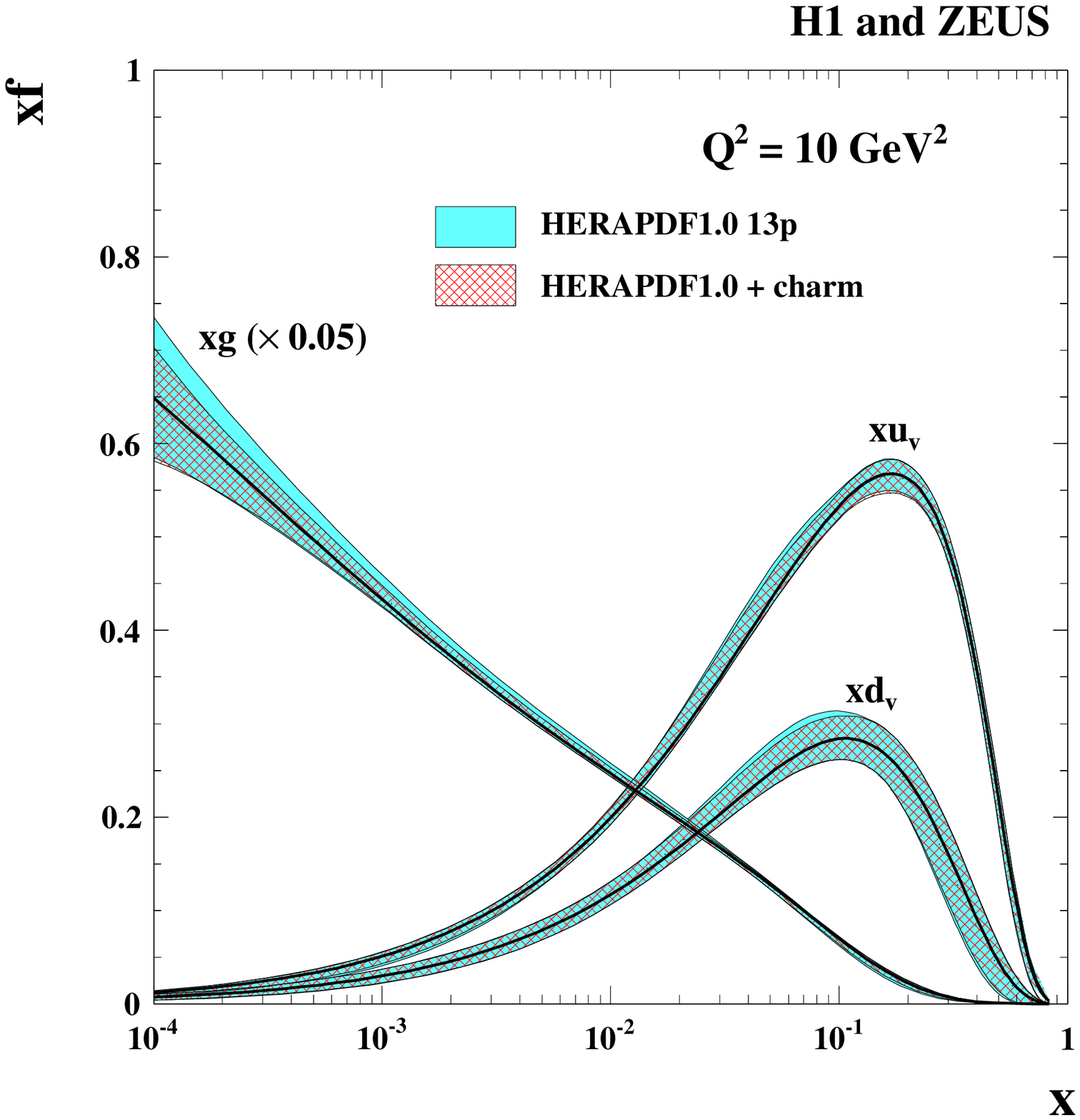,width=.52\textwidth}}
\put(12.,0.){\epsfig{file=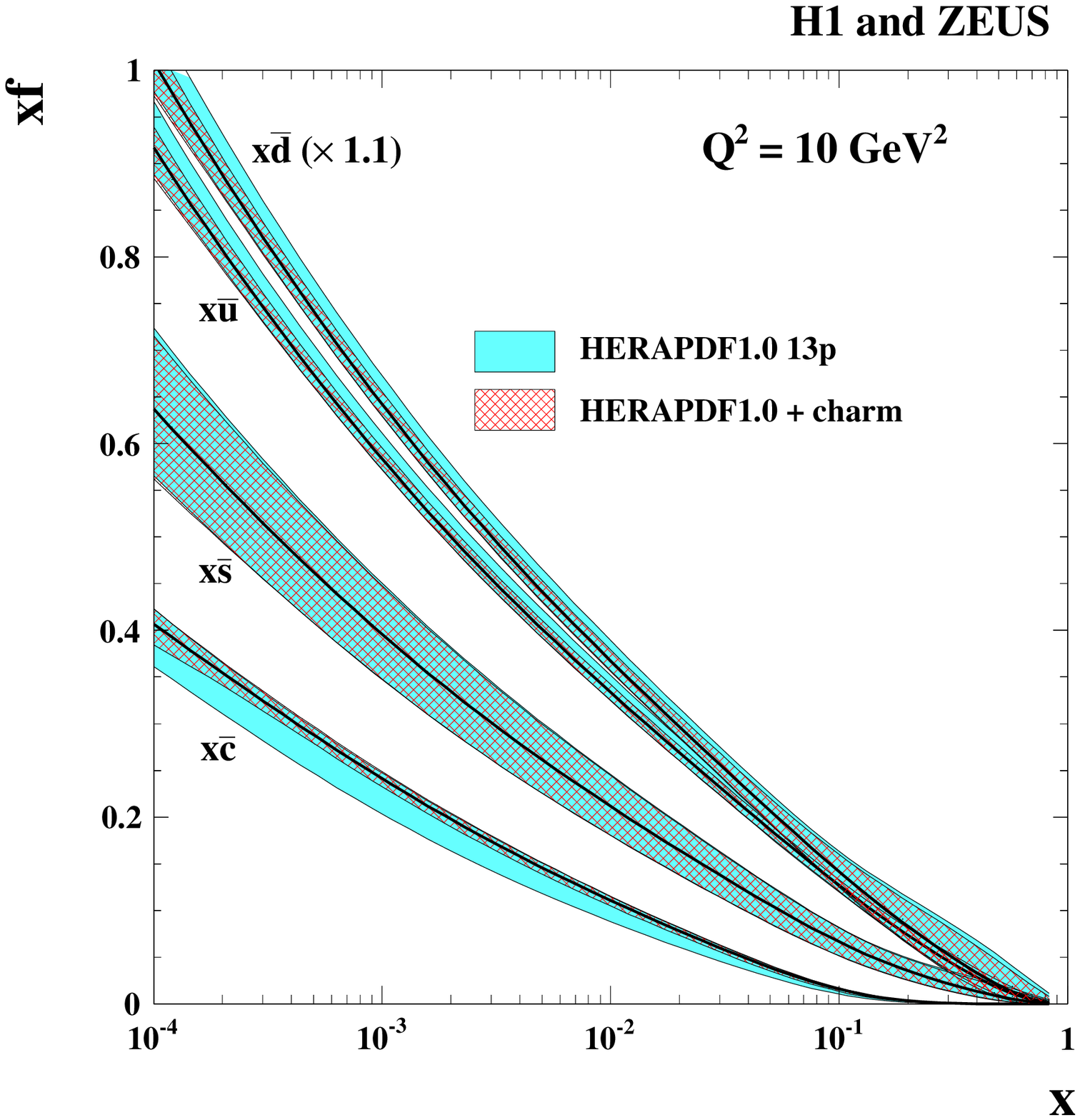,width=.52\textwidth}}
\put(6.2,10.2){\large (a)}
\put(18.2,10.2){\large (b)}
\end{picture}
\caption{Parton density functions $x\cdot f(x,Q^2)$ with $f=g,u_v,d_v,\overline{u},\overline{d},\overline{s},\overline{c}$ for (a) valence quarks and gluon and for (b) sea anti-quarks obtained from the combined QCD analysis of the inclusive DIS data 
and \red (dark shaded bands) in the RT optimised scheme as a function of $x$ at $Q^2=10$ GeV$^2$. For comparison the results of the QCD analysis of the inclusive DIS data only are also shown (light shaded bands). The gluon distribution function is scaled by a factor $0.05$ and the $x\overline{d}$ distribution function is scaled by a factor $1.1$ for better visibility.
}
\label{fig:pdf1} 
\end{sidewaysfigure}

In figure~\ref{fig:pdf1} the PDFs from a 13 parameter fit using the inclusive HERA-I data only 
are compared with the corresponding PDFs when including the combined charm data in the fit.
For both of these fits the RT optimised VFNS is used.
The total PDF uncertainties include the parameterisation and model uncertainties as described in section \ref{sect:qcdana} except for the uncertainties due to \mct, which is treated as follows: 
in the fit based solely on the inclusive data a central value 
of $\mct=1.4$~GeV is used with a variation in the range $1.35<\mct<1.65$~GeV, consistent with the treatment for HERAPDF1.0. 
For the fit including the combined charm cross sections this parameter is set to \mcto~ 
with the corresponding uncertainties as obtained by the charm mass scan for the RT optimised VFNS (table~\ref{tab:mco}).

By comparing the PDF uncertainties obtained from the analysis of the inclusive data only and from the combined analysis of the inclusive and charm data, the following observations can be made: 
\begin{itemize}
\item  the inclusion of \red in the fit does not alter the central PDFs significantly; the central PDFs obtained with the charm data lie well within the uncertainty bands of the PDFs based on the inclusive data only;
\item the uncertainties of the valence quark distribution functions are almost unaffected; 
\item the uncertainty on the gluon distribution function is reduced, mostly due to a reduction in the
parameterisation uncertainty coming from the constraints that the charm data 
put on the gluon through the $\gamma g \to c\overline{c}$ process;
\item the uncertainty on the $x\overline{c}$ distribution function is considerably reduced due to the constrained range of \mct;
\item  the uncertainty on the $x\overline{u}$ distribution function is correspondingly reduced because the inclusive data constrains the sum $x\overline{U}=x\overline{u}+x\overline{c}$;
\item the uncertainty on the $x\overline{d}$ distribution function is also 
reduced because it is constrained to be equal to $x\overline{u}$ at low $x$;
\item the uncertainty on the $x\overline{s}$ distribution function is not reduced because it is dominated by the model uncertainty on the strangeness fraction $f_s$.
\end{itemize}
Similar conclusions hold also for the other schemes discussed in this paper.

\subsection{Measurement of the charm quark mass}

\begin{figure}[hhh]
\center
\epsfig{file=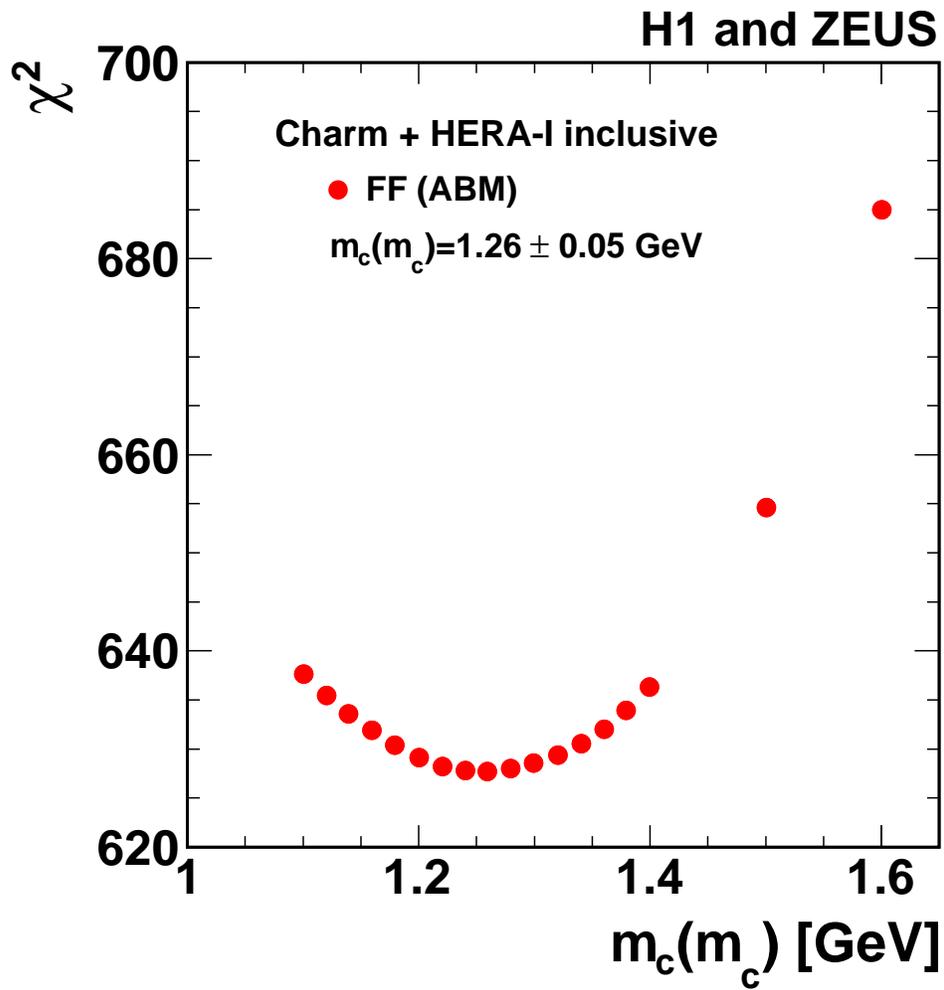,width=01.\textwidth}
\caption{The values of $\chi^2$ for the PDF fit to the combined HERA DIS data including charm measurements 
as a function of the running charm quark mass $m_c(m_c)$. The FFNS ABM scheme is used, where the charm quark 
mass is defined in the $\overline{\text{MS}}$ scheme.}
\label{fig:abm_scan} 
\end{figure}

In the VFNS discussed in the previous section the charm quark mass parameter \mct~does not correspond directly to a physical mass. This is different for the FFNS. 
An NLO QCD analysis is performed in the FFNS of the ABM group~\cite{ABKMMSbar} to determine the $\overline {\text{MS}}$ running charm quark mass $m_c(m_c)$ based on the inclusive neutral and charged current HERA-I DIS data and the charm cross section. For this purpose the coefficient functions as implemented in OPENQCDRAD~\cite{abm11,openqcdrad} are used. The strong coupling constant is evolved with setting the number of active flavours to $n_f=3$, using $\alpha_s^{n_f=3}(M_Z)=0.105$.
The same minimisation procedure as for the VFNS analysis is applied and 
the resulting dependence of the $\chi^2$ values from the QCD fits on the charm quark mass $m_c$ is shown in figure~\ref{fig:abm_scan}. The fit of the parabolic function, defined in equation (\ref{mcfit}), results in a value of 
\begin{equation}
 m_c(m_c)=1.26~\pm~0.05_{\rm exp}\pm 0.03_{\rm mod}\pm 0.02_{\rm param}\pm 0.02_{\alpha_s}\,{\rm GeV}
\label{eqn:mcrun}
\end{equation}
for the running charm mass in NLO.
The errors correspond to the experimental, the model, parameterisation and $\alpha_s$ dependent uncertainties. The same variations of the model and parameterisation assumptions are applied as for the results presented in section \ref{mcvfns} and discussed in section \ref{sect:qcdana}.
The data are well described by the FFNS calculations for $m_c(m_c)=1.26$~GeV with a total $\chi^2=627.7$ for $626$ degrees of freedom. The partial contribution from the charm data is $\chi^2=43.5$ for $47$ data points.
The measured value of the running charm quark mass is consistent with the world average  
of $m_c(m_c)=1.275\pm 0.025$~GeV~\cite{pdg} defined at two-loop QCD, based on lattice calculations and measurements of time-like processes. It also compares well to recent analyses \cite{ABKMMSbar,adlm} of DIS and charm data at NLO and NNLO.

\subsection{Impact of charm data on predictions for {\boldmath $W^\pm$} and {\boldmath $Z$\unboldmath} production at the LHC}

The different series of PDFs obtained from fits to the HERA data by the $\mct$ scanning procedure 
in the different VFNSs are used to calculate cross section 
predictions for $W^{\pm}$ and $Z$ production at the LHC at $\sqrt{s}=7$ TeV. 
These predictions are calculated for each scheme
using the MCFM program~\cite{mcfm} interfaced to APPLGRID~\cite{applgrid}
for $1.2\le \mct \le 1.8$~GeV in $0.1$~GeV steps, except for S-ACOT-$\chi$ for which the range
$1.1\le \mct \le 1.4$~GeV is used.

\begin{figure}[h]
\center
\begin{picture}(16.,19.)
\put(0.,8.8){\epsfig{file=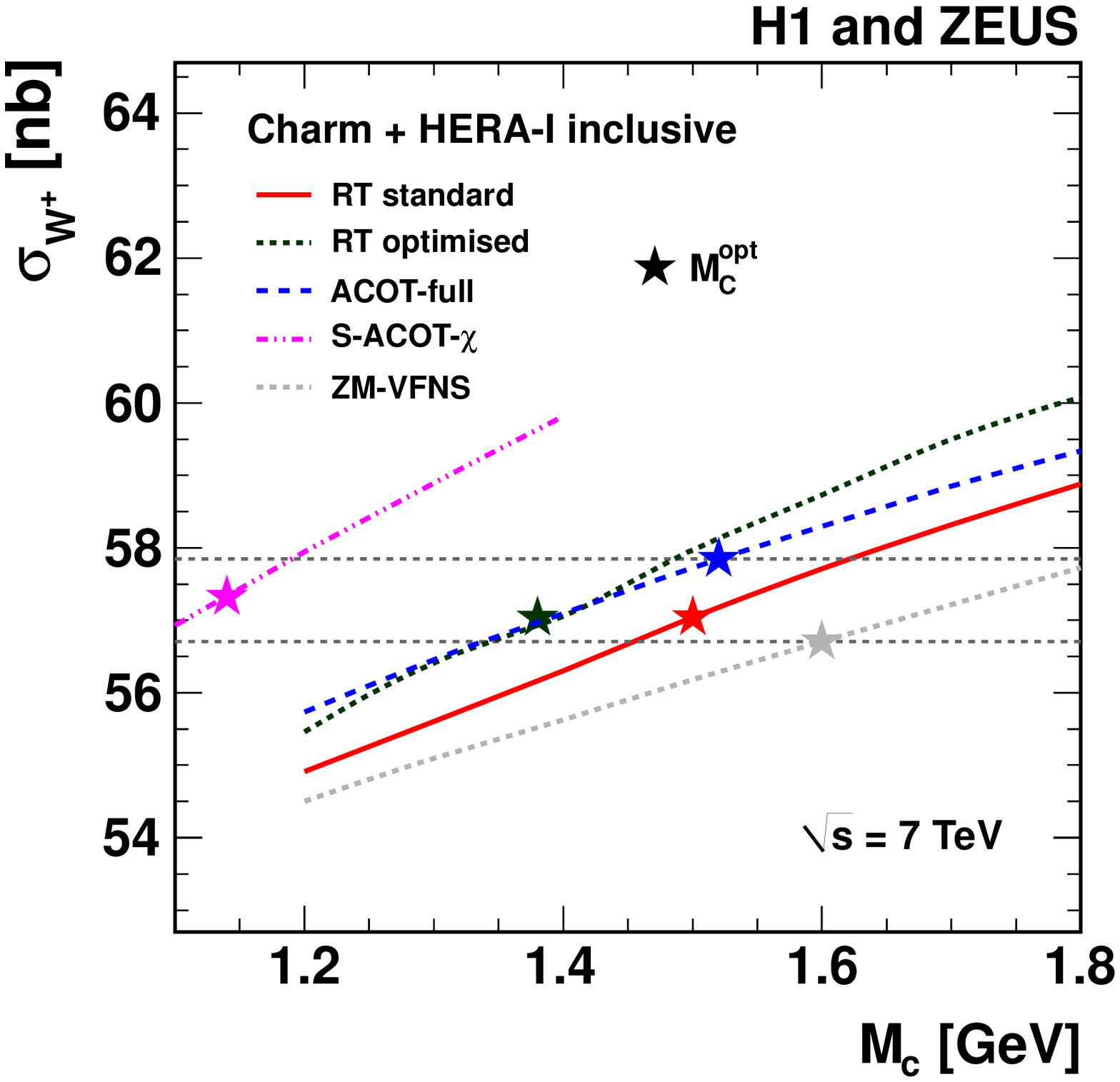,width=0.58\textwidth}}
\put(8.2,8.8){\epsfig{file=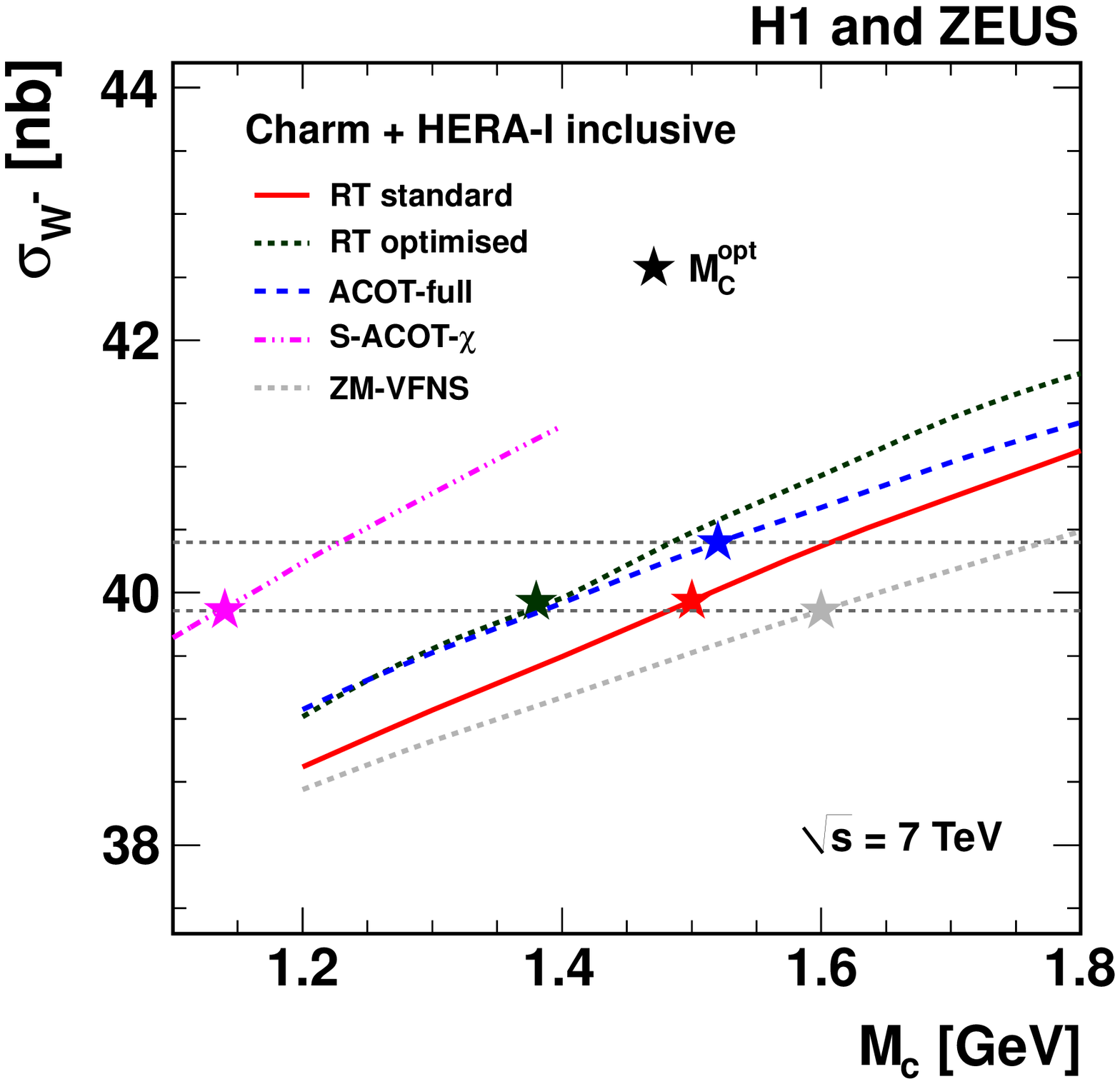,width=0.58\textwidth}}
\put(3.5,0.){\epsfig{file=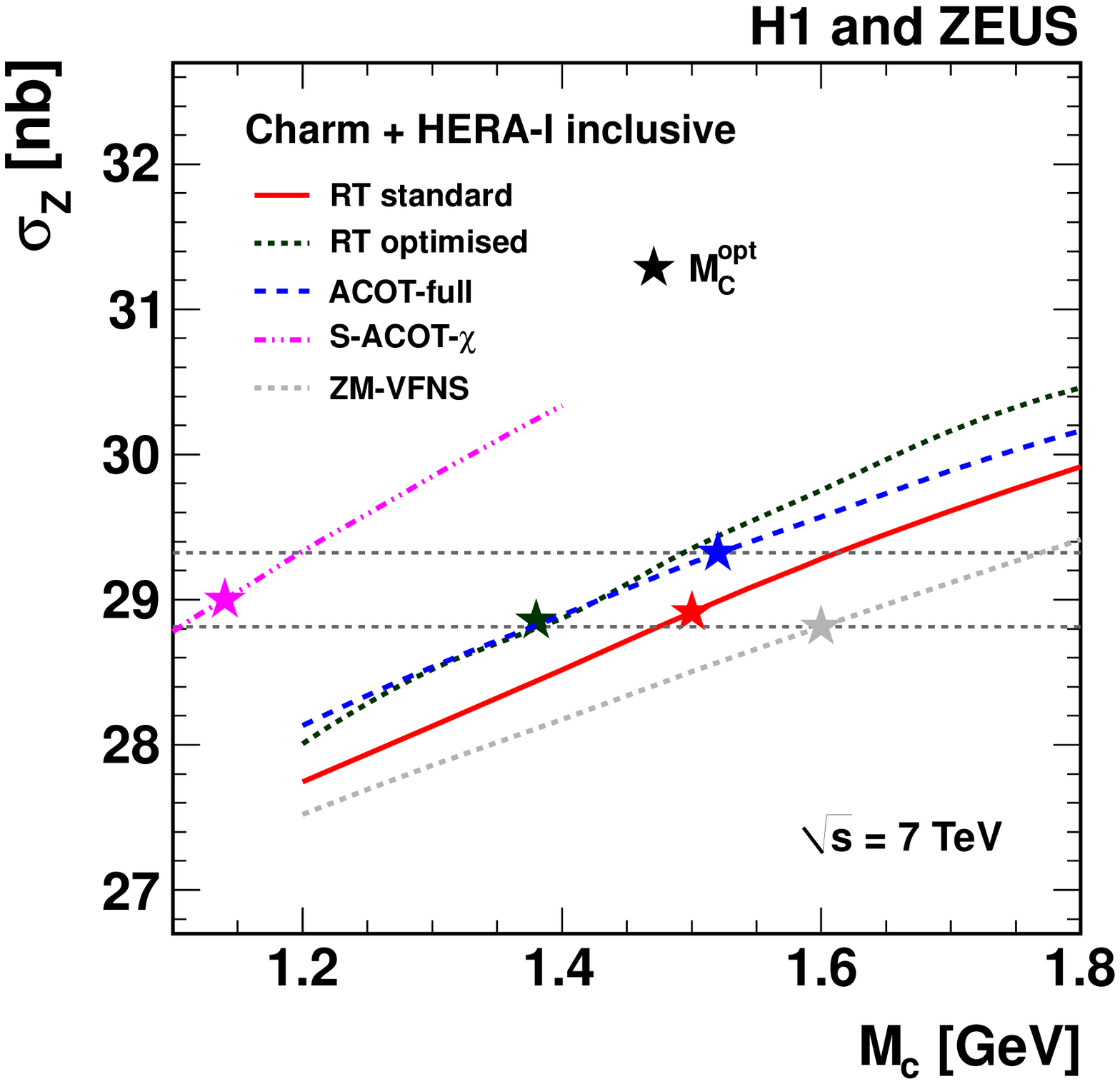,width=0.58\textwidth}}
\put(6.5,15.4){\large (a)}
\put(14.5,15.4){\large (b)}
\put(10.,6.6){\large (c)}
\end{picture}
\caption{NLO predictions for (a) $W^{+}$, (b) $W^{-}$ and (c) $Z$ production cross sections 
at the LHC 
for $\sqrt{s}=7$ TeV as a function of $\mct$ used in the corresponding PDF fit. 
The different lines represent predictions for different implementations of the VFNS. 
The predictions obtained with PDFs 
evaluated with the $\mcto$ values for each scheme are indicated by the stars. 
The horizontal dashed lines show the resulting spread of the predictions 
when choosing $M_c=\mcto$.}
\label{fig:wp_cross_section} 
\end{figure}

The predicted $W^\pm$ and $Z$ production cross sections as a function of $\mct$ for the different implementations 
of the VFNS are shown in figure~\ref{fig:wp_cross_section} and the values for the 
optimal choice $\mcto$ are summarised in table~\ref{tab:WZ_cross_sections}.
For all implementations of VFNS a similar monotonic dependence of the $W^{\pm}$ and $Z$ production 
cross sections on \mct~is observed.
This can be qualitatively understood as follows. 
A higher charm mass leads to stronger suppression of charm near threshold such that more light sea quarks are required to fit the inclusive data. More gluons are also needed to describe the HERA charm data.  The need for more light sea quarks at the initial scale together with the creation of more sea quarks from gluon splitting at higher scales lead to an enhancement of the $W^\pm$ and $Z$ cross sections at the LHC.

There is a significant spread of about $6\%$ between the predictions 
if they are considered for a fixed value of $\mct$, e.g. at \mct$=1.4$ GeV. 
Similarly, the prediction typically 
varies by about $6\%$
when raising $\mct$ from $1.2$ to $1.8$~GeV. However, when using the $\mcto$ for each scheme the spread of predictions is 
reduced to $1.4\%$ for $W^-$, $1.8\%$ for $Z$ and to $2\%$ for $W^+$ production.
 
This indicates that a good description of the HERA charm 
data correlates with a very similar flavour composition of the quark PDFs at 
LHC scales, almost independent of the chosen scheme. The uncertainty on 
the $W^\pm$ and $Z$ cross section predictions due to the choice of the charm mass
can thus be considerably reduced. However, the charm 
mass parameter must be adjusted to a different value for each scheme, 
consistent with the HERA data, in order to achieve this result.

\section{Conclusions}
\label{sect:conclude}
Measurements of open charm production in deep-inelastic $ep$ scattering by the H1 and ZEUS experiments using different charm tagging methods are 
combined, accounting for the systematic correlations. The measurements 
are extrapolated to the full phase space
using an NLO QCD calculation to obtain the reduced charm quark-pair cross sections in the region of photon 
virtualities $2.5\le Q^2 \le 2000$ GeV$^2$. The combined data are compared to QCD predictions in the fixed-flavour-number-scheme and in the general-mass variable-flavour-number-scheme. 
The best description of the data in the whole kinematic range is provided by the NNLO fixed-flavour-number-scheme prediction of the ABM group.
Some of the NLO general-mass variable-flavour-number-scheme predictions significantly underestimate the charm production cross section at low $Q^2$, which is 
improved at NNLO. 

Using the combined charm cross sections together with the combined HERA inclusive DIS data, an NLO 
QCD analysis is performed based on different implementations of the 
variable-flavour-number-scheme. For each scheme, an optimal 
value of the charm mass parameter, \mcto, is determined. These values show a 
sizeable spread. All schemes are found to describe the data well, as long as the charm mass parameter is taken at the 
corresponding optimal value. The use of \mcto~ and its uncertainties in the QCD analysis significantly reduces the parton density uncertainties, mainly for the sea quark contributions from charm, down and up quarks.

The QCD analysis is also performed in the fixed-flavour-number-scheme at NLO using the $\overline{\rm MS}$ running mass definition. The running charm quark mass is determined as $ m_c(m_c)=1.26~\pm~0.05_{\rm exp.}\pm 0.03_{\rm mod}\pm 0.02_{\rm param}\pm 0.02_{\alpha_s}$ GeV. This value agrees well with the world average based on lattice calculations and on measurements of time-like processes.

The PDFs obtained from the corresponding QCD analyses using different \mct~  
are used to predict $W^\pm$ and $Z$ production cross-sections at the LHC. A sizeable spread in 
the predictions is observed, when the charm mass parameter \mct~ is varied between 1.2 and 1.8 GeV, 
or when different schemes are considered at fixed value of \mct. The spread is significantly 
reduced when the optimal value of \mct~ is used for each scheme. 
\section*{Acknowledgements}

We are grateful to the HERA machine group whose outstanding
efforts have made these experiments possible.
We appreciate the contributions to the construction and maintenance of the H1 and ZEUS detectors of many people who are not listed as authors.
We thank our funding agencies for financial 
support, the DESY technical staff for continuous assistance and the 
DESY directorate for their support and for the hospitality they 
extended to the non-DESY members of the collaborations. 

We also would like to thank S. Alekhin, P. Nadolsky, J. Rojo and R. Thorne for providing predictions and for fruitful discussions.
We thank M. Karnevsky for contributions to the $W^\pm$ and $Z$ cross section calculations.
\noindent
\begin{flushleft}

\end{flushleft}
\newpage
\begin{table}[h]
\begin{center}
\begin{tabular}{|rl|l|rcr|r|r|}\hline 
       \multicolumn{2}{|l|}{Data set}               & Tagging method 
     & \multicolumn{3}{|c|}{$Q^2$ range}       & $N$ & ${\cal L}$  \\ 
     &                                      &                  &
                                           \multicolumn{3}{|c|}{[GeV$^2$]}
                                           &  & [pb$^{-1}$]  \\ \hline
1&H1 VTX~\cite{h1ltt_hera2}    &Inclusive track lifetime&  $5$ &--& $2000$                & $29$ & $245$ \\
2&H1 $D^*$ HERA-I~\cite{h1dstar_hera1}      &$D^{*+}$    &  $2$ &--& $100$                  & $17$ & $47$\\
3&H1 $D^*$ HERA-II~\cite{h1dstar_hera2}     &$D^{*+} $     &  $5$ &--& $100$                  &  $25$ & $348$\\
4&H1 $D^*$ HERA-II~\cite{h1dstarhighQ2}     &$D^{*+}$     &  $100$ &--& $1000$            &  $6$ & $35$1 \\
5&ZEUS $D^*$ (96-97)~\cite{zd97}           &$D^{*+}$      & $1$ &--& $200$                  & $2$1 & $37$ \\
6&ZEUS $D^*$ (98-00)~\cite{zd00}           &$D^{*+}$      &  $1.5$ &--& $100$0            & $31$ & $82$ \\
7&ZEUS $D^0$~\cite{zd0dp}                  &$D^{0,{\rm no}D^{*+}}$ &$ 5$ &--& $1000$    & $9$  & $134$\\
8&ZEUS $D^+$~\cite{zd0dp}                  &$D^+$         & $5$ &--& $1000$               & $9$   & $134$\\
9&ZEUS $\mu$~\cite{zmu}                    &$\mu$         & $20$ &--& $10000$          & $8$  &  $12$6  \\ \hline
\end{tabular}
\end{center}
\caption{Data sets used in the combination. For each data set the charm tagging method, the $Q^2$ range, the number of cross section measurements $N$ and the integrated luminosity
  ${\cal L}$ are given. The data set with the $D^{0,{\rm no}D^{*+}}$ tagging method
is based on an analysis of \dzero~mesons not originating from detectable $D^{*+}$
decays. Charge conjugate modes are always implied.
}
\label{t:sets}
\end{table}
\begin{table}[h]
\begin{center}
\begin{tabular}{|c|c|c|l|} \hline
  $\hat{s}$ range       &  $\alpha_K (D^*)$  &$\alpha_K (\rm g.s.)$   &  Measurement \\ \hline
  $\hat{s}\leq\hat{s}_1$            &  $6.1 \pm 0.9$     & $4.6\pm 0.7$          &\cite{h1frag} $D^*$, DIS, no-jet sample \\ \hline
  $\hat{s}_1<\hat{s} \leq\hat{s}_2$ &  $3.3 \pm 0.4$     & $3.0\pm 0.3$          &\cite{h1frag} $D^*$, DIS, jet sample   \\ \hline
  $\hat{s}>\hat{s}_2$                        &  $2.67 \pm 0.31$ &  $2.19\pm  0.24$    &\cite{zeusdmesons} $D^*$ jet photoproduction       \\ \hline
\end{tabular}
\end{center}
\caption{The $\alpha_K$ parameters used for the longitudinal fragmentation into
   $D^*$ mesons and in ground state (g.s.) charmed hadrons.  The first column shows the $\hat{s}$ range in which a particular value of  $\alpha_K$  is
  used, with $\hat{s}_1=70\pm40$~GeV$^2$ and $\hat{s}_2=324$~GeV$^2$.
   The variations of $\alpha_K$ are given in the second and third column. 
The parameter $\hat{s}_2$ is not varied, since the 
corresponding uncertainty is already covered by the $\alpha_K$ variations.
\label{t:kart}}
\end{table}
\begin{table}[h]
\begin{center}
\begin{tabular}{|l|c|}  \hline 
$f(c\rightarrow D^{*+})$ & $0.2287 \pm 0.0056$  \\ \hline
$f(c\rightarrow D^+)$   &  $0.2256 \pm 0.0077$\\    \hline
$f(c\rightarrow D^{0, {\rm not}D^{*+}})$  &$0.409 \pm 0.014$\\ \hline
$B(c\rightarrow \mu)$ &  $0.096 \pm 0.004$\\ \hline
\end{tabular}
\end{center}
\caption{Charm fragmentation fractions to charmed mesons and the charm branching fraction to muons. 
\label{t:ff}}
\end{table}
\unitlength1cm
\begin{table}
\renewcommand{\arraystretch}{1.2}
\centering
\begin{tabular}[t]{|c|c|c|r|r|r|r|r|}\hline
$Q^2 [\mathrm{GeV}^2]$  & $x$ & $y$ & \red & $\delta_{unc} [\%]$ & $\delta_{cor} [\%]$ & $\delta_{proced} [\%]$ & $\delta_{tot} [\%]$\\ \hline
  $   2.5 $ & $   0.00003 $ & $   0.824 $ & $ 0.1126 $ & $    14.0  $ & $   10.9  $ & $    0.3 $ & $   17.8$ \\ 
  $   2.5 $ & $   0.00007 $ & $   0.353 $ & $ 0.1068 $ & $     9.0  $ & $    9.9  $ & $    0.2 $ & $   13.4$ \\ 
  $   2.5 $ & $   0.00013 $ & $   0.190 $ & $ 0.0889 $ & $     10.0  $ & $    9.1  $ & $    2.2 $ & $   13.7$ \\ 
  $   2.5 $ & $   0.00018 $ & $   0.137 $ & $ 0.0907 $ & $     9.5  $ & $    8.3  $ & $    1.4 $ & $   12.7$ \\ 
  $   2.5 $ & $   0.00035 $ & $   0.071 $ & $ 0.0560 $ & $     8.7  $ & $    8.2  $ & $    0.0 $ & $   11.9$ \\
  $   5. $ & $   0.00007 $ & $   0.706 $ & $ 0.1466 $ & $    15.6  $ & $    10.0  $ & $    0.2 $ & $   18.5$ \\
  $   5 $ & $   0.00018 $ & $   0.274 $ & $ 0.1495 $ & $     8.4  $ & $    6.8  $ & $    1.1 $ & $   10.8$ \\
  $   5 $ & $   0.00035 $ & $   0.141 $ & $ 0.1151 $ & $     7.1  $ & $    6.7  $ & $    0.6 $ & $    9.8$ \\
  $   5 $ & $   0.00100 $ & $   0.049 $ & $ 0.0803 $ & $     9.2  $ & $    8.2  $ & $    0.6 $ & $   12.4$ \\
  $   7 $ & $   0.00013 $ & $   0.532 $ & $ 0.2142 $ & $     8.1  $ & $    8.0  $ & $    0.2 $ & $   11.4$ \\
  $   7 $ & $   0.00018 $ & $   0.384 $ & $ 0.1909 $ & $    10.2  $ & $    8.5  $ & $    2.1 $ & $   13.4$ \\
  $   7 $ & $   0.00030 $ & $   0.231 $ & $ 0.1689 $ & $     4.6  $ & $    6.3  $ & $    0.4 $ & $    7.8$ \\
  $   7 $ & $   0.00050 $ & $   0.138 $ & $ 0.1553 $ & $     4.3  $ & $    5.9  $ & $    0.6 $ & $    7.3$ \\
   $  7 $ & $   0.00080 $ & $   0.086 $ & $ 0.1156 $ & $     7.2  $ & $    6.0  $ & $    0.7 $ & $    9.4$ \\
  $   7 $ & $   0.00160 $ & $   0.043 $ & $ 0.0925 $ & $     6.4  $ & $    7.6  $ & $    0.6 $ & $    9.9$ \\
  $  12 $ & $   0.00022 $ & $   0.539 $ & $ 0.2983 $ & $     8.4  $ & $    7.2  $ & $    0.1 $ & $   11.1$ \\
  $  12 $ & $   0.00032 $ & $   0.371 $ & $ 0.2852 $ & $     4.7  $ & $    6.5  $ & $    0.6 $ & $    8.1$ \\
  $  12 $ & $   0.00050 $ & $   0.237 $ & $ 0.2342 $ & $     4.3  $ & $    5.1  $ & $    0.5 $ & $    6.6$ \\
  $  12 $ & $   0.00080 $ & $   0.148 $ & $ 0.1771 $ & $     3.8  $ & $    5.7  $ & $    0.1 $ & $    6.9$ \\
  $  12 $ & $   0.00150 $ & $   0.079 $ & $ 0.1413 $ & $     5.5  $ & $    6.8  $ & $    0.1 $ & $    8.7$ \\
  $  12 $ & $   0.00300 $ & $   0.040 $ & $ 0.1028 $ & $     6.1  $ & $    8.0  $ & $    0.2 $ & $   10.1$ \\
  $  18 $ & $   0.00035 $ & $   0.508 $ & $ 0.3093 $ & $     9.2  $ & $    6.5  $ & $    1.0 $ & $   11.3$ \\
  $  18 $ & $   0.00050 $ & $   0.356 $ & $ 0.2766 $ & $     4.7  $ & $    7.0  $ & $    0.5 $ & $    8.4$ \\
  $  18 $ & $   0.00080 $ & $   0.222 $ & $ 0.2637 $ & $     3.8  $ & $    4.6  $ & $    0.6 $ & $    6.1$ \\
  $  18 $ & $   0.00135 $ & $   0.132 $ & $ 0.2009 $ & $     3.3  $ & $    5.2  $ & $    0.0 $ & $    6.2$ \\
  $  18 $ & $   0.00250 $ & $   0.071 $ & $ 0.1576 $ & $     3.5  $ & $    5.7  $ & $    0.1 $ & $    6.7$ \\
  $  18 $ & $   0.00450 $ & $   0.040 $ & $ 0.1349 $ & $     5.8  $ & $    8.0  $ & $    1.4 $ & $    10.0$ \\
\hline
\end{tabular}
\caption{The averaged reduced cross section of charm production, \red, as obtained from the
combination of H1 and ZEUS measurements. The values of the cross section are presented together 
with uncorrelated ($\delta_{unc}$) correlated ($\delta_{cor}$) and procedural ($\delta_{proced}$) uncertainties. 
The total uncertainty ($\delta_{tot}$) is obtained by adding the correlated, uncorrelated and procedural errors in quadrature.}
\label{tab:result}
\end{table}
\unitlength1cm
\begin{table}
\renewcommand{\arraystretch}{1.2}
\centering
\begin{tabular}[t]{|c|c|c|r|r|r|r|r|}\hline
$Q^2 [\mathrm{GeV}^2]$  & $x$ & $y$ & \red & $\delta_{unc} [\%]$ & $\delta_{cor} [\%]$ & $\delta_{proced} [\%]$ & $\delta_{tot} [\%]$\\ \hline

  $  32 $ & $   0.00060 $ & $   0.527 $ & $ 0.4119 $ & $    15.1  $ & $    5.7  $ & $    0.1 $ & $   16.2$ \\
  $  32 $ & $   0.00080 $ & $   0.395 $ & $ 0.3527 $ & $     4.3  $ & $    5.3  $ & $    0.3 $ & $    6.9$ \\
  $  32 $ & $   0.00140 $ & $   0.226 $ & $ 0.2767 $ & $     3.9  $ & $    4.2  $ & $    0.4 $ & $    5.8$ \\
  $  32 $ & $   0.00240 $ & $   0.132 $ & $ 0.2035 $ & $     4.8  $ & $    4.9  $ & $    0.3 $ & $    6.9$ \\
  $  32 $ & $   0.00320 $ & $   0.099 $ & $ 0.1942 $ & $     7.1  $ & $    5.6  $ & $    0.3 $ & $    9.0$ \\
  $  32 $ & $   0.00550 $ & $   0.058 $ & $ 0.1487 $ & $     6.9  $ & $    6.0  $ & $    0.4 $ & $    9.1$ \\
  $  32 $ & $   0.00800 $ & $   0.040 $ & $ 0.1027 $ & $    10.7  $ & $    8.3  $ & $    0.4 $ & $   13.5$ \\
  $  60 $ & $   0.00140 $ & $   0.424 $ & $ 0.3218 $ & $     6.1  $ & $    5.4  $ & $    1.4 $ & $    8.3$ \\
 $   60 $ & $   0.00200 $ & $   0.296 $ & $ 0.3387 $ & $     4.3  $ & $    3.7  $ & $    0.4 $ & $    5.7$ \\
  $  60 $ & $   0.00320 $ & $   0.185 $ & $ 0.2721 $ & $     4.7  $ & $    3.9  $ & $    0.4 $ & $    6.1$ \\
  $  60 $ & $   0.00500 $ & $   0.119 $ & $ 0.1975 $ & $     4.7  $ & $    4.9  $ & $    0.3 $ & $    6.8$ \\
  $  60 $ & $   0.00800 $ & $   0.074 $ & $ 0.1456 $ & $    12.0  $ & $    5.2  $ & $    0.6 $ & $   13.1$ \\
  $  60 $ & $   0.01500 $ & $   0.040 $ & $ 0.1008 $ & $    10.6  $ & $    6.4  $ & $    0.8 $ & $   12.4$ \\
 $  120 $ & $   0.00200 $ & $   0.593 $ & $ 0.3450 $ & $     7.1  $ & $    5.2  $ & $    0.6 $ & $    8.8$ \\
 $  120 $ & $   0.00320 $ & $   0.371 $ & $ 0.2432 $ & $    15.9  $ & $    4.0  $ & $    2.1 $ & $   16.5$ \\
 $  120 $ & $   0.00550 $ & $   0.216 $ & $ 0.2260 $ & $     5.2  $ & $    4.5  $ & $    0.6 $ & $    6.9$ \\
 $  120 $ & $   0.01000 $ & $   0.119 $ & $ 0.1590 $ & $     6.6  $ & $    5.4  $ & $    0.8 $ & $    8.6$ \\
 $  120 $ & $   0.02500 $ & $   0.047 $ & $ 0.0866 $ & $    13.7  $ & $    6.8  $ & $    1.2 $ & $   15.3$ \\
 $  200 $ & $   0.00500 $ & $   0.395 $ & $ 0.2439 $ & $     8.1  $ & $    5.7  $ & $    0.7 $ & $    9.9$ \\
 $  200 $ & $   0.01300 $ & $   0.152 $ & $ 0.1659 $ & $     6.7  $ & $    4.8  $ & $    0.4 $ & $    8.3$ \\
 $  350 $ & $   0.01000 $ & $   0.346 $ & $ 0.2250 $ & $     8.8  $ & $    5.0  $ & $    4.1 $ & $   10.9$ \\
 $  350 $ & $   0.02500 $ & $   0.138 $ & $ 0.1016 $ & $    11.2  $ & $    5.8  $ & $    5.1 $ & $   13.6$ \\
 $  650 $ & $   0.01300 $ & $   0.494 $ & $ 0.2004 $ & $    11.1  $ & $    7.2  $ & $    1.1 $ & $   13.3$ \\
 $  650 $ & $   0.03200 $ & $   0.201 $ & $ 0.0939 $ & $    12.4  $ & $   10.6  $ & $    0.9 $ & $   16.4$ \\
$  2000 $ & $   0.05000 $ & $   0.395 $ & $ 0.0622 $ & $    27.7  $ & $   14.4  $ & $    1.7 $ & $   31.2$ \\ \hline
\end{tabular}
\captcont{continued}

\end{table}
\begin{table}
\centering
\footnotesize
\begin{tabular}[t]{|c|l|l|c|c|}\hline
source & data sets & name & shift $[\sigma]$ & Reduction factor $[\%]$\\ \hline
 $\delta_1$ & 1 &H1 vertex resolution        & -0.1 &   94 \\
 $\delta_2$ & 1--4 &H1 CJC efficiency        & -0.3 &   82 \\
 $\delta_3$ & 1 &H1 CST efficiency           &  0.0 &   98 \\
 $\delta_4$ & 1 &B multiplicity              & -0.3 &   96 \\
 $\delta_5$ & 1--9 &{\bf\boldmath $c$ longitudinal fragmentation}  & -0.9 &   84 \\
 $\delta_6$ & 1, 3, 4 & photoproduction background                 &  0.2 &   94 \\
 $\delta_7$ & 1& $D^+$ multiplicity                &  0.0 &   99 \\
 $\delta_8$ & 1& $D^0$ multiplicity                &  0.0 &   99 \\
 $\delta_9$ & 1& $D_s$ multiplicity                &  0.1 &   98 \\ 
 $\delta_{10}$ & 1&$b$ fragmentation               & 0.0 &   100 \\
 $\delta_{11}$ & 1&H1 VTX model: $x$-reweighting   & -0.4 &   95 \\
 $\delta_{12}$ & 1&H1 VTX model: $p_T$-reweighting &  0.3 &   74 \\
 $\delta_{13}$ & 1&H1 VTX model: $\eta(c)$-reweighting  & -0.3 &   87 \\
 $\delta_{14}$ & 1&H1 VTX $uds$-background              & 0.0 &   53 \\
 $\delta_{15}$ & 1&H1 VTX $\phi$ of $c$-quark           &  0.2 &   90 \\
 $\delta_{16}$ & 1&H1 hadronic energy scale             & -0.1 &   89 \\
 $\delta_{17}$ & 1&H1 VTX $F_2$ normalisation           & -0.2 &   97 \\
 $\delta_{18}$ & 3, 4& H1 Primary vertex fit            &  0.1 &   99 \\
 $\delta_{19}$ & 2--4&H1 electron energy                &  0.6 &   69 \\
 $\delta_{20}$ & 2--4&H1 electron polar angle           &  0.3 &   77 \\
 $\delta_{21}$ & 3, 4 &H1 luminosity (HERA-II)          & -0.9 &   80 \\
 $\delta_{22}$ & 3, 4&H1 trigger efficiency (HERA-II)   & -0.3 &   98 \\
 $\delta_{23}$ & 3, 4&H1 fragmentation model in MC      & -0.1 &   89 \\
 $\delta_{24}$ & 2--7 & $BR(D^*\to K \pi \pi)$          &  0.1 &   98 \\ 
 $\delta_{25}$ & 2--6 &{\bf\boldmath $f(c \to D^*)$}                       &  0.1 &   94. \\
 $\delta_{26}$ & 2, 3& H1 efficiency using alternative MC model            &  0.4 &   73 \\
 $\delta_{27}$ & 2--9 &{\bf\boldmath NLO, $m_c$}                           &  0.5 &   72 \\
 $\delta_{28}$ & 2--9 &{\bf NLO, scale}                                    & -1.2 &   66 \\
 $\delta_{29}$ & 2--9 &{\bf\boldmath $c$ transverse fragmentation    }     & -0.2 &   78 \\
 $\delta_{30}$ & 2--9 &{\bf NLO, PDF                    }                  &  0.2 &   97 \\
 $\delta_{31}$ & 2--9 &{\bf NLO, \boldmath$\alpha_s(M_Z)$        }         & -0.2 &   95 \\
 $\delta_{32}$ & 2 &H1 luminosity (1998-2000)                              & -0.1 &   97 \\
 $\delta_{33}$ & 2 &H1 trigger efficiency (HERA-I)                         & -0.2 &   95 \\
 $\delta_{34}$ & 2 &H1 MC alternative fragmentation                        & -0.1 &   70 \\
 $\delta_{35}$ & 9 &ZEUS $\mu$: B/RMUON efficiency                         & -0.1 &   92 \\
 $\delta_{36}$ & 9 &ZEUS $\mu$: FMUON efficiency                           &  0.2 &   97 \\
 $\delta_{37}$ & 9&ZEUS $\mu$: energy scale                                &  0.0 &   85 \\
 $\delta_{38}$ & 9&ZEUS $\mu$: $P_{T}^{\rm miss}$ calibration              &  0.0 &   72 \\
 $\delta_{39}$ & 9&ZEUS $\mu$: hadronic resolution                         &  0.6 &   71 \\
 $\delta_{40}$ & 9&ZEUS $\mu$: IP resolution                               & -0.2 &   97 \\
 $\delta_{41}$ & 9&ZEUS $\mu$: MC model                                    &  0.1 &   86 \\
 $\delta_{42}$ & 9&{\boldmath $B(c \to \mu)$}                              &  0.1 &   97 \\
 $\delta_{43}$ & 7, 8 &ZEUS lifetime significance                          &  0.5 &   52 \\
 $\delta_{44}$ & 7 &{\boldmath$f(c \to D^0)$}                              &  0.3 &   97 \\
 $\delta_{45}$ & 8& {\boldmath $f(c \to D^+) \times BR(D^+ \rightarrow K\pi\pi)$} & -0.6 &   91 \\
 $\delta_{46}$ & 7--9 & ZEUS luminosity (2005)                             & -0.1 &   95 \\
 $\delta_{47}$ & 5 &ZEUS luminosity (1996-1997)                            &  0.4 &   96 \\
 $\delta_{48}$ & 6 &ZEUS luminosity (1998-2000)                            &  0.3 &   90 \\ \hline
\end{tabular}

\caption{Sources of bin-to-bin correlated systematic uncertainties
  considered in the combination. For each source the shifts in units
  of standard deviations $\sigma$ and the reduction factor of the uncertainty values are
  given. The systematic sources corresponding to the extrapolation
  uncertainties are highlighted in bold font. The second column shows
  the data sets (see table \ref{t:sets}) affected by each particular source.}
\label{tab:systematic_shifts}

\end{table}
\begin{sidewaystable}[p]
\centering
\linespread{1.3}
  \normalsize 
{\small  \begin{tabular}[h]{|l|l|l|c|c|c|c|c|c|l|}
    \hline
   Theory           & Scheme        & Ref.               & $F_{2(L)}$    & $m_c$      & Massive  & Massless  & $\alpha_s(m_Z)$ & Scale & Included  \\
                    &               &                    &  def.        &  [GeV]&
                    $(Q^2\lsim m_c^2)$  & $(Q^2\gg m_c^2)$  & ($n_f=5$)  &            & charm data  \\
   \hline
  MSTW08 NLO        & RT standard   &\cite{rt_std}       & $F_{2(L)}^c$  & $1.4$ (pole)        & ${\cal O}(\alpha_s^2)$ & ${\cal O}(\alpha_s)$    & $0.12108$    &$Q$ &\cite{h196,zd97,h1f2c,zd00,h1vertex05,h1ltt_hera1,zeusdmesons}  \\ 
  MSTW08 NNLO       &               &                    &              &            & approx.-${\cal O}(\alpha_s^3)$ & ${\cal O}(\alpha_s^2)$ & $0.11707$ & &     \\ 
  MSTW08 NLO (opt.) & RT optimised  &\cite{rt_opt}       &              &            & ${\cal O}(\alpha_s^2)$ & ${\cal O}(\alpha_s)$    & $0.12108$       & &     \\ 
  MSTW08 NNLO (opt.)&               &                    &              &            & approx.-${\cal O}(\alpha_s^3)$ & ${\cal O}(\alpha_s^2)$ & $0.11707$ & &     \\
  \hline 
 HERAPDF1.5 NLO        & RT standard   &\cite{ref:HERAPDF1.5}       & $F_{2(L)}^c$  & $1.4$ (pole)        & ${\cal O}(\alpha_s^2)$ & ${\cal O}(\alpha_s)$    & $0.1176$    &$Q$ &HERA inclusive DIS only  \\  
\hline 
  NNPDF2.1 FONLL A  & FONLL A       &\cite{fonllb_and_c} & n.a.         & $\sqrt{2}$  & ${\cal O}(\alpha_s)$   & ${\cal O}(\alpha_s)$    & $0.119$ & $Q$ & \cite{zd97,h1f2c,zd00,zd0dp,zmu,h1dstarhighQ2,h1dstar_hera2}     \\
  NNPDF2.1 FONLL B  & FONLL B       &                    & $F_{2(L)}^c$  & $\sqrt{2}$ (pole)    & ${\cal O}(\alpha_s^2)$ & ${\cal O}(\alpha_s)$  & & &        \\
  NNPDF2.1 FONLL C  & FONLL C       &                    & $F_{2(L)}^c$  &  $\sqrt{2}$ (pole)    & ${\cal O}(\alpha_s^2)$ & ${\cal O}(\alpha_s^2)$  &           & &  \\
  \hline 
  CT10  NLO         & S-ACOT-$\chi$ &\cite{ct10f3}         & n.a.         & $1.3$         & ${\cal O}(\alpha_s)$   & ${\cal O}(\alpha_s)$    & $0.118$ & $\sqrt{Q^2 + m_c^2}$ & \cite{zd97,h1f2c,zd00,h1vertex05,h1ltt_hera1}   \\
  CT10  NNLO (prel.)&               &\cite{ct12nnlo}  & $F_{2(L)}^{c\bar c}$ & $1.3$ (pole)   & ${\cal O}(\alpha_s^2)$ & ${\cal O}(\alpha_s^2)$  &       &        &    \\
  \hline
  ABKM09 NLO        & FFNS  &\cite{abkm}  & $F_{2(L)}^{c\bar c}$  & $1.18$ ($\overline{\text {MS}}$) & ${\cal O}(\alpha_s^2)$ & - & 0.1135 & $\sqrt{Q^2+4m_c^2}$ & for mass optimisation only   \\
  ABKM09 NNLO       &      &               &                     & & approx.-${\cal O}(\alpha_s^3)$ & - &   &     &         \\
  \hline
   \end{tabular}}
\linespread{1.0}
  \caption{Calculations from different theory groups as shown in 
figures~\ref{fig:combined_mstw}-\ref{fig:combined_abkm}. The table
shows the heavy flavour scheme used and the corresponding reference,
the respective $F_{2(L)}$ definition (section \ref{sect:theory}), 
the value and type of charm mass used (equation (\ref{eq:mrunpol})), 
the order in $\alpha_S$ of the massive and massless
parts of the calculation, the value of $\alpha_s$, the renormalisation and factorisation scale, and which HERA charm data were included in the corresponding PDF fit. The distinction between the two possible $F_{2(L)}$ definitions is not applicable (n.a.) for ${\cal O}(\alpha_s)$ calculations.}    
\label{tab:theories}
\end{sidewaystable}
\begin{table}[!ht]
\centering
\linespread{1.3}
  \normalsize 
  \begin{tabular}[h]{|c|c|c|c|}
    \hline
   scheme   & \mcto\,& $\chi^2/n_{\rm dof}$ & $\chi^2/{n_{\rm dp}}$ \\
            &  [GeV]   &    $\sigma^{NC,CC}_{\rm red}$+\red          & \red   \\
  \hline
  \hline 
      RT standard   & $1.50 \pm 0.06_{\text{exp}} \pm 0.06_{\text{mod}} \pm 0.01_{\text{param}} \pm 0.003_{\alpha_s}$ & $630.7/626$& $49.0/47$\\
      RT optimised  & $1.38 \pm 0.05_{\text{exp}} \pm 0.03_{\text{mod}} \pm 0.01_{\text{param}} \pm 0.01_{\alpha_s}$ & $623.8/626$& $45.8/47$\\
      ACOT-full     & $1.52 \pm 0.05_{\text{exp}} \pm 0.12_{\text{mod}} \pm 0.01_{\text{param}} \pm 0.06_{\alpha_s}$ & $607.3/626$& $53.3/47$\\
      S-ACOT-$\chi$ & $1.15 \pm 0.04_{\text{exp}} \pm 0.01_{\text{mod}} \pm 0.01_{\text{param}} \pm 0.02_{\alpha_s}$ & $613.3/626$& $50.3/47$\\
      ZM-VFNS        & $1.60 \pm 0.05_{\text{exp}} \pm 0.03_{\text{mod}} \pm 0.05_{\text{param}} \pm 0.01_{\alpha_s}$ & $631.7/626$& $55.3/47$\\ [2pt]
  \hline
   \end{tabular}
\linespread{1.0}
  \caption{ The values of the charm mass parameter \mcto\ as determined from the \mct\ scans in different heavy flavour schemes. The uncertainties of the minimisation procedure 
are denoted as `exp', the model and parameterisation uncertainties are represented by `mod' and `param', respectively. Also the uncertainty due to 
$\alpha_s$ variation is listed.
The corresponding global and partial $\chi^2$ are presented per degrees of freedom $n_{\text {dof}}$ and per number of data points $n_{\text {dp}}$, respectively. }    
\label{tab:mco} 
\end{table}
\begin{table}[!ht]
\centering
\linespread{1.3}
  \normalsize 
  \begin{tabular}[h]{|c|c|c|c|}
    \hline
   scheme   & $\sigma_{Z}$ [nb] & $\sigma_{W^{+}}$ [nb] & $\sigma_{W^{-}}$ [nb] \\
            &                  &                      &  \\
  \hline
  \hline 
      RT standard   &  $28.91 \pm 0.30 $&$ 57.04 \pm 0.55 $&$ 39.94 \pm 0.35 $\\
      RT optimised  &  $28.85 \pm 0.24 $&$ 57.03 \pm 0.45 $&$ 39.93 \pm 0.27 $\\
      ACOT-full     &  $29.32 \pm 0.42 $&$ 57.84 \pm 0.74 $&$ 40.39 \pm 0.47 $\\
      S-ACOT-$\chi$ &  $29.00 \pm 0.22 $&$ 57.32 \pm 0.42 $&$ 39.86 \pm 0.24 $\\
      ZM-VFNS        &  $28.81 \pm 0.24 $&$ 56.71 \pm 0.40 $&$ 39.86 \pm 0.25 $\\ [2pt]
  \hline
   \end{tabular}
\linespread{1.0}
  \caption{NLO VFNS predictions for $Z/W^\pm$ cross sections at the LHC for $\sqrt{s}=7$ TeV using the MCFM program. The calculations are based on the PDF sets extracted in the corresponding schemes from the HERA data using \mcto~ for the charm quark mass parameter.
The listed cross section uncertainties correspond to the uncertainties on \mcto~only.}    
\label{tab:WZ_cross_sections} 
\end{table}
\newpage

\newpage

\newpage

\newpage

\newpage

\newpage

\newpage

\newpage

\newpage

\newpage

\newpage
\newpage

\newpage

\unitlength1cm
\end{document}